\documentclass[aps,prd,superscriptaddress]{revtex4-2}
\usepackage{xspace,graphicx,verbatim}
\usepackage{amsmath,amssymb}
\usepackage[squaren]{SIunits}
\usepackage{hyperref}
\usepackage{rotating}

\hypersetup{
    colorlinks=true, % false: boxed links; true: colored links
    linkcolor=red,   % color of internal links
    citecolor=blue,  % color of links to bibliography
    filecolor=blue,  % color of file links
    urlcolor=blue    % color of external links
}

\newcommand{\me}{\text{e}}
\newcommand{\dift}[1]{\frac{\mathrm{d}#1}{\mathrm{d}t}}
\addunit\perh{\hour^{-1}}

\newcommand{\cresta}{Cresta et al.\ \cite{cresta21}\xspace}
\newcommand{\simon}{Simon et al.\ \cite{simon16}\xspace}
\newcommand{\quirouette}{Quirouette et al.\ \cite{quirouette23}\xspace}
\newcommand{\prior}{\mathcal{P}_\text{prior}}
\newcommand{\Like}{\mathcal{L}}
\newcommand{\LikeNR}{\Like_\text{NR}}
\newcommand{\LikeED}{\Like_\text{ED}}
\newcommand{\sdata}{\{\text{data}\}}
\newcommand{\MM}{\text{MM}}
\newcommand{\Cdata}{\ensuremath{C^\text{data}}\xspace}

\newcommand{\RNS}{\ensuremath{\text{RNS}}\xspace}
\newcommand{\rna}{_\text{RNA}}
\newcommand{\tinf}{t_\text{inf}}

\addunit\SIN{SIN}
\newcommand{\tcid}{\ensuremath{\text{TCID}_{50}}\xspace}
\newcommand{\Vinoc}{\ensuremath{V_\text{inoc}}\xspace}
\newcommand{\Vvir}{\ensuremath{V_\text{vir}}\xspace}
\newcommand{\Dil}{\ensuremath{\mathcal{D}}\xspace}

\newcommand{\col}{\ensuremath{\text{col}}\xspace}

\addunit\IV{IV}
\newcommand{\Vol}{s}
\newcommand{\Ncells}{\ensuremath{N_\text{cells}}\xspace}
\newcommand{\ssigma}{(\beta \Ncells/\Vol)/c}
\newcommand{\burst}{\ensuremath{\mathcal{B}}\xspace}
\newcommand{\PVext}{\ensuremath{\mathcal{P}_{V \to \, \text{Extinction}}}\xspace}
\newcommand{\PVest}{\ensuremath{\mathcal{P}_{V \to \, \text{Establishment}}}\xspace}
\newcommand{\PVtoI}{\mathcal{P}_{V \to \, I }}
\newcommand{\pset}{\vec{\pi}}
\newcommand{\pinf}{p_\text{inf}}
\newcommand{\mapVOmc}{2\,259}

\begin{document}

\title{Does the random nature of cell-virus interactions during in vitro infections affect $\tcid$ measurements and parameter estimation by mathematical models?}

\author{Christian Quirouette}
\affiliation{Department of Physics, Toronto Metropolitan University, Toronto, Canada}

\author{Risavarshni Thevakumaran}
\altaffiliation{Present address: Department of Biomedical Engineering, McGill University, Montr\'eal, Canada}
\affiliation{Department of Electrical, Computer and Biomedical Engineering, Toronto Metropolitan University, Toronto, Canada}

\author{Kyosuke Adachi}
\affiliation{Interdisciplinary Theoretical and Mathematical Sciences (iTHEMS), RIKEN, Wako, Japan}
\affiliation{Nonequilibrium Physics of Living Matter RIKEN Hakubi Research Team, RIKEN Center for Biosystems Dynamics Research, Kobe, Japan}

\author{Catherine A.\ A.\ Beauchemin}
\email[Corresponding author: ]{cbeau@torontomu.ca}
\affiliation{Department of Physics, Toronto Metropolitan University, Toronto, Canada}
\affiliation{Interdisciplinary Theoretical and Mathematical Sciences (iTHEMS), RIKEN, Wako, Japan}

\date{\today}

\begin{abstract} % About 361 words
Endpoint dilution (\tcid) assays cannot count the number of infectious virions (IVs), and instead are limited to counting the number of specific infections caused by the sample (SIN). The latter quantity depends not only on whether virions are infectious, but also on the cells and the experimental conditions under which they interact. These interactions are random and controlled by infection parameters, such as the rates at which IVs lose infectivity, enter cells, or fail to replicate following cell entry.
Here, we simulate stochastic \tcid assays to determine how the random number of infected wells relates to the infection parameters and the number of IVs in a sample. We introduce a new parameter estimation method based on the likelihood of observing a given \tcid assay outcome given the model-predicted number of IVs in the sample.
We then successively evaluate how parameter estimation is affected by the use of: (1) the new likelihood function versus the typical assumption of Gaussian-distributed measurement errors; (2) IV versus SIN to represent virus in the mathematical model; and (3) a stochastic model rather than a deterministic ordinary differential equation (ODE) model to simulate the course of a virus infection.
Unlike previous methods, the new likelihood intuitively handles measurements beyond the limits of detection, and results in non-Gaussian distributions for certain measurements.
Expressing virus using units of IV makes it possible to impose physical constraints (one IV cannot infect more than one cell, the number of IVs cannot exceed the number of viral RNA, etc.), and yields more biologically useful parameters, e.g.\ the likelihood of mutation emergence depends on the total number of IVs, not SIN, produced.
Using a stochastic rather than an ODE model had a negligible impact on parameter estimation, because in vitro infections are intentionally inoculated with enough IVs to guarantee infection will establish, resulting in negligible stochasticity. Instead, we show that the variability observed between triplicate in vitro virus infections is consistent with the level of stochasticity introduced by the \tcid assay, which can be reduced through better assay design.
The new modelling and parameter estimation framework introduced herein offers several important improvements over current methods and should be widely adopted.
\end{abstract}

\maketitle

\section{Introduction}

Experimental measurements of the virus concentration in the supernatant of an infected cell culture or in samples from an infected host are the primary quantity through which infection progression is monitored and evaluated. Fold-reduction in quantities such as peak or total viral yield under antiviral therapy is a common endpoint to quantify antiviral efficacy \cite{hayden99,hayden00,barroso05,devincenzo15,hayden18,stevens18,devincenzo22,hammond22}, and the degree to which viral yield correlates with disease severity or transmissibility is often investigated \cite{devincenzo10,houben10,takeyama12,feikin15,granados17,goikhman20,mckay20,creager23,kieran24}. But total and peak viral yields are the result of many different aspects of virus replication and cell-virus interactions, such as the virus' affinity for a cell's surface receptors, the rate of virus production by an infected cell and the length of time for which an infected cell produces virus.

Mathematical models (MMs) enable us to bridge the gap between intra- and inter-cellular events (causes) and experimental observations (effects). They play a critical role in elucidating and challenging our understanding of the mechanisms behind the establishment and progression of virus infections in vitro and within hosts. MMs achieve this through a mathematical description of cell-virus interaction mechanisms, each characterized by one or more parameters controlling changes in mathematical variables which directly correspond or can be related to experimental measurements. We infer the most likely value of the MM parameters, and thus characterize the relative contribution of individual mechanisms, by matching MM-predicted variables to their corresponding experimental measurements. It is therefore critical to correctly identify and mathematically represent the relationship between experimental measurements and their corresponding MM variables.

Infectious virions, the causative agents of the infection, are often one of the variables in MMs of viral infections. Experimentally, it is not the concentration of infectious virions in a sample that is measured, but rather a quantity related to the number of cells the sample is expected to infect. These experimental measures are obtained via two main assays: the plaque (or focus) forming assay and the endpoint dilution (or \tcid) assay \cite{spearman1908,karber31,reed38,dulbecco52,dulbecco55,wulff12,mistry18,cresta21}. In a typical plaque (or focus) forming assay, the virus sample is serially diluted to an extent where ideally, within one well containing about one million cells, only about 30 cells will become infected so that only about 30 plaque forming units (PFUs) are counted. This is to avoid a count that is too low and therefore highly variable (1 to 5 plaques) or too high resulting in overlapped or merged plaques which would make counting ambiguous. For its part, the \tcid assay relies on estimating the dilution of the virus sample at which it would infect 50\% of the cell culture wells and thus contain \emph{on the order of} a single infection-causing dose. Both assays assume or rather define that one infection-causing dose inevitably causes an infection. There is, therefore, a fundamental difference between a sample's experimentally measured concentration of infection-causing doses (e.g., PFU, \tcid), and its concentration of infectious virions as described by MMs, wherein one infectious virion is capable of but not assured to cause an infection.

Given that the endpoints of these assays are based on a small number of infection-causing doses, one might wonder to what extent stochastic effects, e.g.\ the random chance that one infectious virion fails to initiate an infection, affect the accuracy of the infection-causing dose estimated from the assays. Several past works using stochastic mathematical models (SMs) have evaluated the likelihood that an infection, starting from one or very few infectious virions or infected cells, will establish or will become extinct, and found this likelihood varies as a function of the SM parameters \cite{conway11,pearson11,conway13,heldt15,yan16,conway18,czuppon21,quirouette23}. The relationship between the SM's infectious virions and the experimentally measured infection-causing doses is therefore nontrivial. Importantly, it depends on the very quantities the MM aims to determine from these experimental measurements: its parameters.

\begin{figure}
\begin{center}
\includegraphics[width=0.4\linewidth]{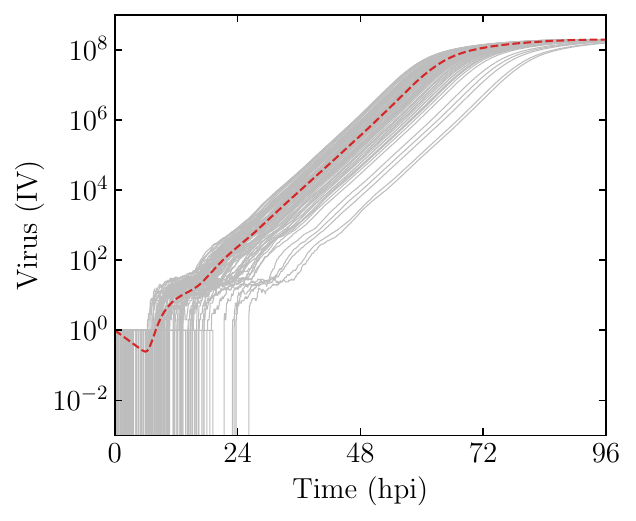}
\end{center}
\caption{%
\textbf{Random events can cause a time shift of infection curves.} %
Infectious virion time courses for 100 SM simulations (grey solid lines) compared to the ODE solution (red dashed line), for an infection initiated with a single infectious virion. Infection parameters were taken from \simon with some adaptations for the MMs used herein as described in Methods, except for $\beta = \unit{10^{-6}}{\milli\litre/(cell\cdot\hour)}$ which was reduced to enhance the variability in the curve timing.%
}
\label{tshift}
\end{figure}

Typical MMs of virus infections are deterministic, mean-field, ordinary differential equation (ODE) models. For an infection initiated with very few infectious virions or cells, the random nature of the first few infection events will determine whether the infection will fail or successfully establish. The timing of these early infection events will set the timing of the entire infection curve \cite{sazonov20,morris24}. Thereafter, as the number of infectious virions and cells continue to increase, the behaviour becomes deterministic, following the shape of the ODE-predicted infectious virus time course. The stochasticity, it would seem, translates the ODE-predicted curve to earlier or later times, but does not affect its overall shape, as illustrated in Figure \ref{tshift}. As such, for a given set of MM parameters, a stochastic infection will yield a family of time-offset curves which, if analyzed with an ODE model, could result in different parameter estimates.

In this work, we use a SM largely identical to that previously introduced \cite{quirouette23} to study how its parameters affect the relationship between the actual number of infectious virions represented by the SM and the number of infection-causing doses that would be experimentally measured via an endpoint dilution (\tcid) assay. Taking this relationship into consideration, we introduce a novel framework with which to interpret experimental infectious dose measurements and to relate them to their corresponding MM-predicted variables. We then compare results of the typical parameter estimation approach using an ODE model against individual components of the new framework. We explore the implications of our findings on interpretations of present and past results.

%%%%%%%%%%%%%%%%%%%%%%%%%%%%%%%%%%%%%%%%%%%%%%%%%%%%%%%%%%%%%%%%%%%%%%%%%%%%%%%%
%%%%%%%%%%%%%%%%%%%%%%%%%%%%%%%%%%%%%%%%%%%%%%%%%%%%%%%%%%%%%%%%%%%%%%%%%%%%%%%%
%%%%%%%%%%%%%%%%%%%%%%%%%%%%%%%%%%%%%%%%%%%%%%%%%%%%%%%%%%%%%%%%%%%%%%%%%%%%%%%%

\section{Results}

\subsection{Relationship between infectious virions and experimental measurements}

\begin{figure}
\begin{center}
\includegraphics[width=\linewidth]{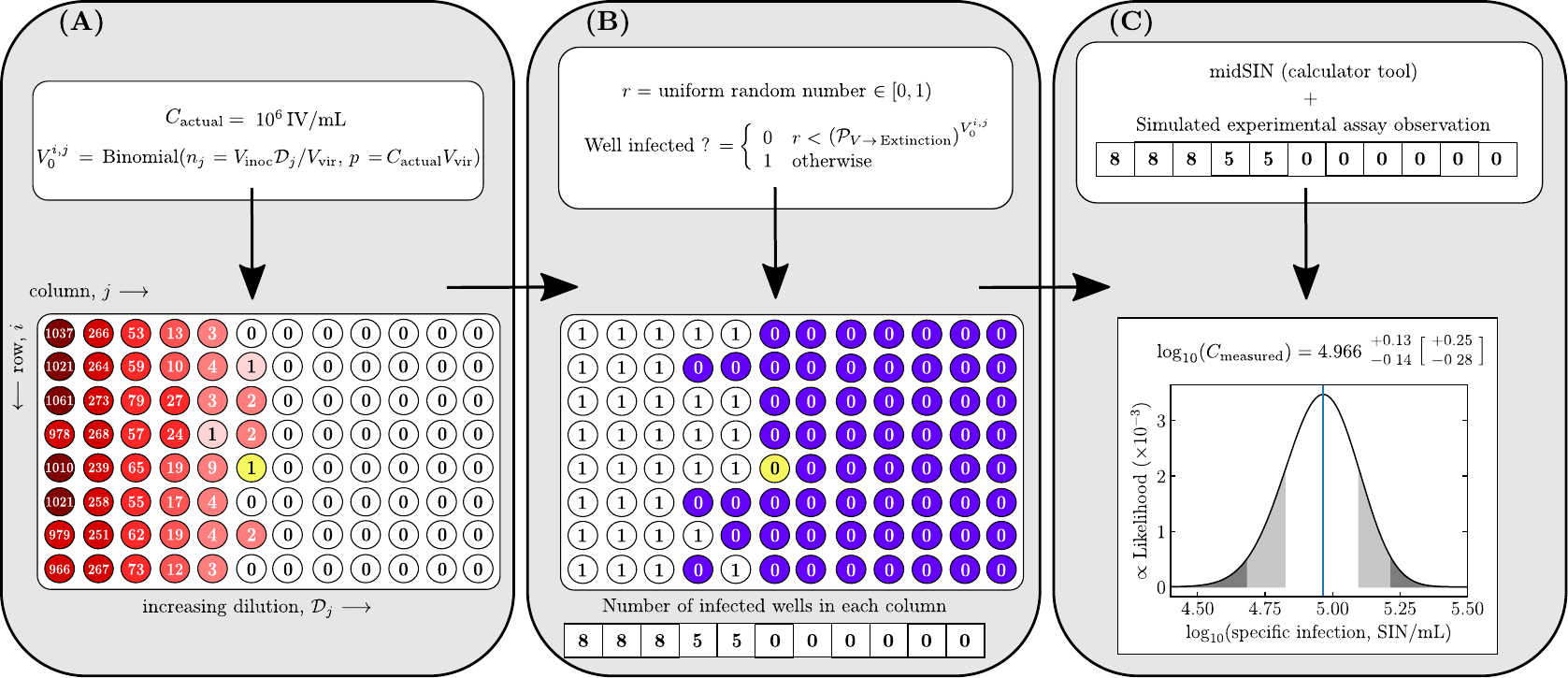}
\end{center}
\caption{%
\textbf{Stochastic scheme to simulate an ED assay experiment.} %
(A) The random number of infectious virions (IV) inoculated into each well ($V_0^{i,j}$) is drawn from a Binomial distribution, based on the known virus sample concentration ($C_\text{actual}$ in \IV/\milli\litre), the dilution factor in that well's column ($\Dil_j\in(0,1]$), the volume of a single virion ($\Vvir$), and the total inoculum volume placed in each well ($\Vinoc$ = diluted sample plus dilution medium).
(B) The random outcome for each well, i.e.\ whether infection took place or not (1 or 0, respectively), is determined based on the number of IV received by the well ($V_0^{i,j}$) and the SM-predicted likelihood that infection will fail ($\PVext$). From this, the overall experimental assay outcome, namely the number of infected wells observed in each dilution column, is computed.
(C) The midSIN calculator tool \cite{cresta21} is used to estimate the sample's most likely $\log_{10}$ infectious dose concentration, in units of \SIN/\milli\litre, based on the experimental outcome observed in (B).
The well highlighted in yellow (row 5, column 6) illustrates how a well can receive one infectious virion and yet fail to cause an infection.
}
\label{algo}
\end{figure}

A \tcid endpoint dilution (ED) assay is usually carried out in a cell culture plate consisting of a number of wells organized into rows and columns. Let us consider an example ED experiment in which the first 11 columns of the 96-well plate receive increasing dilutions of the virus sample, all 8 rows within a column receive the same dilution and thus serve as replicates, and the last ($12^\text{th}$) column is reserved for the sample-free control. Figure \ref{algo}A illustrates one possible random outcome for the number of infectious virions that would be received by each well of this simulated experiment given a virus sample with a known infectious virion (IV) concentration ($C_\text{actual}=\unit{10^6}{\IV/\milli\litre}$). the dilution factor for each column ($\Dil_1=10^{-2}$, $\Dil_2=10^{-2.6}$, $\Dil_3=10^{-3.2}$, ..., $\Dil_{11}=10^{-8}$), and the total volume of inoculum placed in each well ($\Vinoc = \unit{0.1}{\milli\litre}$). See Methods for additional details.

The experimental readout of an ED assay is the number of wells that became infected at each dilution as a result of the number of infectious virions each well received given the virus sample's unknown concentration (Fig.\ \ref{algo}B). Past work using SMs suggests that, over a wide range of viruses studied to date (influenza \cite{yan16}, HIV \cite{pearson11,conway13,conway18}, SARS-CoV-2 \cite{czuppon21,quirouette23}) and for realistic infection conditions and parameters, an infection initiated with one or even a few infectious virions can randomly become extinct, i.e.\ fail to take hold or spread significantly. This is more likely to occur when the inoculum consists of only a few infectious virions or the virus' replicative fitness is such that each infected cell barely infects one other cell (basic reproductive number $R_0 \sim 1$). This means that an ED well might fail to become infected despite having received one or even a few infectious virions.

\begin{figure}
\begin{center}
\includegraphics[width=\linewidth]{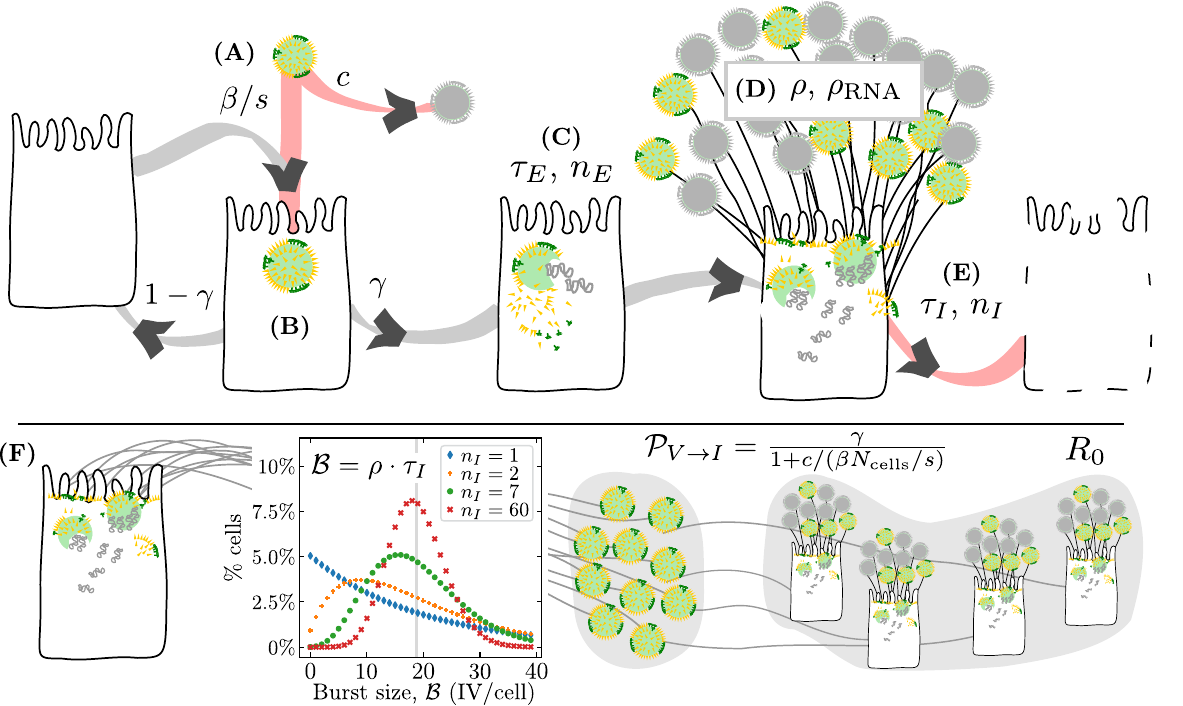}
\end{center}
\caption{%
\textbf{Key parameters and terms of the ODE and stochastic models}.
The ODE and stochastic models describe the virus infection of a population of $N_\text{cells}$ cells, bathed in a supernatant of volume $\Vol$ (see Methods). (A) Infectious virions (IV) in the supernatant are lost due to loss of infectivity at rate $c$ or irreversible entry into cells at rate $\beta/\Vol$. (B) When one IV is lost to cell entry, it either successfully infects it, resulting in $\gamma$ infected cell per IV entry, or it does not ($1-\gamma$). (C) There is a delay of $\tau_E\pm (\tau_E/\sqrt{n_E})$ hours following a cell's infection (eclipse phase) before it begins releasing IV progeny. (D) Thereafter, it releases infectious virions into the supernatant at rate $\rho$, and total (infectious + non-infectious) virion progeny at rate $\rho\rna$. (E) Virus release will persist for $\tau_I\pm (\tau_I/\sqrt{n_I})$ hours (infectious phase) after which the cell undergoes apoptosis. (F) An infected cell will produce on average $\burst=\rho\cdot\tau_I$ IV over its infectious lifespan, but the actual number varies randomly cell-to-cell based on parameter $n_I$ (graph modified from \cite{quirouette23} where $\burst = \unit{19}{IV/cell}$). Each produced IV has a probability $\PVtoI$ to cause the infection of another cell. The basic reproductive number, $R_0 = \burst \cdot \PVtoI$, corresponds to the average number of cells one infected cell will infect. In the illustration, the infected cell produced 10 $\IV$ progeny (based on $\burst$, $n_I$) which went on (based on $\PVtoI$) to productively infect 4 cells ($R_0=4$).
}
\label{figMM}
\end{figure}

The work herein will make use of the virus infection stochastic model, and a related ODE model, largely identical to those introduced and studied previously in \quirouette. One additional equation was added to each MM herein to represent total virions ($V\rna$) corresponding to the number of viral RNA copies as measured by qRT-PCR in units of viral RNA (vRNA). Details of the two MMs are provided in the Methods, and the key parameters and expressions they share, are illustrated in Figure \ref{figMM}. In \cite{quirouette23}, we derived the probability that an infection initiated with a single infectious virion (IV) will become extinct (\PVext), and its dependence on the SM's parameters, namely
\begin{align}
\PVext &=
\left[1-\frac{\gamma}{1+c/(\beta N_\text{cells}/\Vol)}\right] +
\frac{\gamma}{1+c/(\beta N_\text{cells}/\Vol)} \left[\frac{\burst\,(1-\PVext)}{n_I}+1\right]^{-n_I}
\label{PVext}
\end{align}
for which only a numerical solution exists. The term $\PVtoI=\gamma/[1+c/(\beta\Ncells/\Vol)]$ appears twice and corresponds to the probability that one $\IV$ productively infects one cell, with $\gamma\in(0,1]$ the probability that an $\IV$'s entry into a cell will result in the cell's productive infection. When re-written as $[\gamma\beta\Ncells/\Vol]/[\beta\Ncells/\Vol + c]$, it is the ratio of the rate at which target cells are lost to infection ($\gamma\beta\Ncells/\Vol$) divided by the rate at which infectious virions are lost due to cell entry ($\beta\Ncells/\Vol$) plus loss of infectivity ($c$). The first term in Eqn.\ \eqref{PVext}, $1-\PVtoI$, is the probability that the $\IV$ fails to infect a cell, and the second is the probability that it infects a cell ($\PVtoI$) times the probability that its $\IV$ progeny (or its progeny's progeny) fails to do the same. The latter depends on the random total number of $\IV$ produced by each infected cell, which varies cell-to-cell following a negative binomial distribution whose mean ($\burst = \rho \cdot \tau_I$) is the product of the infected cell's $\IV$ production rate ($\rho$) and the duration of its infectious period ($\tau_I$), and its variance ($\burst+\burst^2/n_I$) can be controlled independently of its mean via parameter $n_I$ (Figure \ref{figMM}F). Smaller $n_I$ lead to larger cell-to-cell variations in the number of IV produced. The basic reproductive number ($R_0 = \PVtoI \cdot \burst$) is the product of the mean number of $\IV$ produced per cell ($\burst$) times the probability that one $\IV$ infects one cell ($\PVtoI$). Eqn.\ \eqref{PVext} shows that $\PVext$ depends differently on $\gamma$, $n_I$, and on the parameter groupings $[\rho\cdot\tau_I]$ and $[(\beta\Ncells/\Vol)/c]$.

In Figure \ref{algo}B, we use $\PVext$ to generate one random instance of the outcome of infection in each well of the ED assay, i.e.\ whether or not infection establishes (1 or 0, respectively), based on the probability that a well having received $V_0^{i,j}$ IVs will go extinct, $(\PVext)^{V_0^{i,j}}$. In the example, this process resulted in (8,8,8,5,5,0,0,0,0,0,0) infected wells out of 8 wells in total at each of the 11 dilutions. From such an ED outcome, i.e.\ the number of wells infected at each dilution, the Reed-Muench \cite{reed38} or Spearman-K\"arber \cite{spearman1908,karber31} method is typically used to estimate the number of \tcid/\milli\litre\ in the sample, wherein \unit{1}{\tcid} is the dose at which a sample would be expected to infect 50\% (4/8) of wells. Alternatively, from the same ED outcome, the online tool midSIN can be used to estimate the most likely $\log_{10}$ number of infections the sample can cause per unit volume, expressed in units of specific infections or \SIN/\milli\litre\ \cite{cresta21}. Figure \ref{algo}C presents the midSIN-estimated posterior distribution for the $\log_{10}$ \SIN/\milli\litre\ concentration in the example sample, given the observed ED outcome. Here, the sample with an actual concentration of $C_\text{actual}=\unit{10^6}{\IV/\milli\litre}$ would have been measured experimentally to contain $C_\text{measured}=\unit{10^5}{\SIN/\milli\litre}$. This means that, for the SM parameters employed here, $\sim$10 infectious virions on average are required to cause one infection (\unit{10}{\IV/\SIN}).

\begin{figure}
\begin{center}
\includegraphics[width=0.8\linewidth]{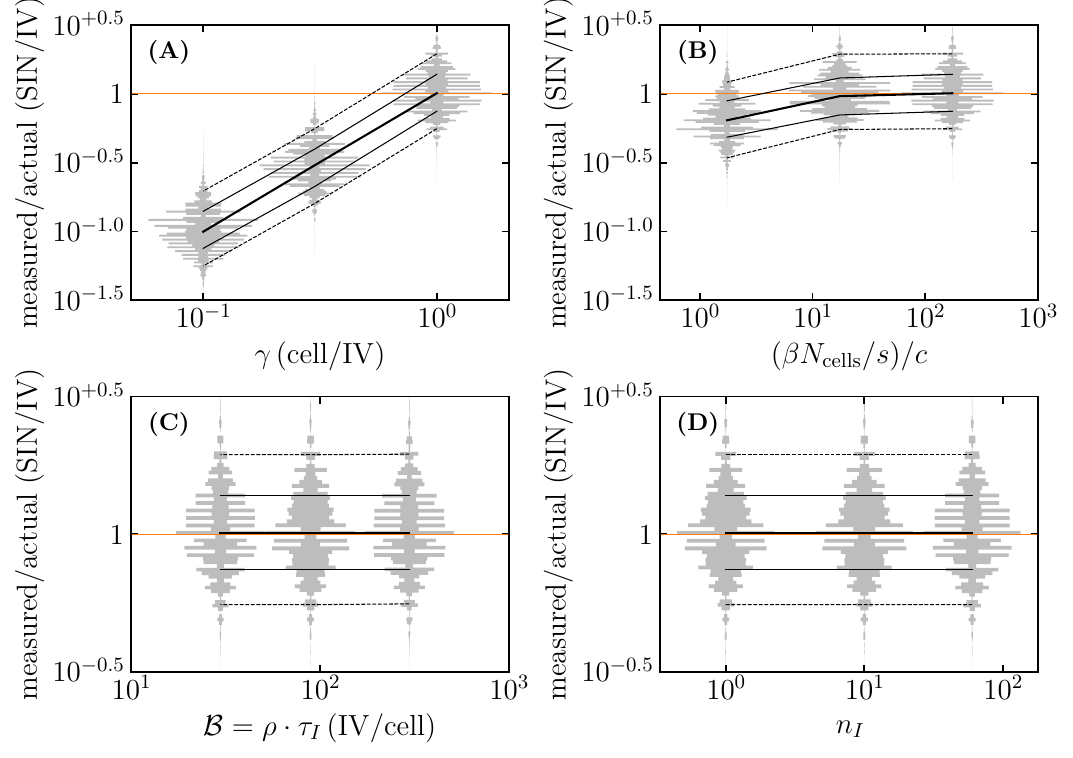}
\end{center}
\caption{%
\textbf{The effect of infection parameters on experimentally measured infection-causing doses.} %
Violin plot (vertically stacked histograms) for the ratio of experimentally \emph{measured} virus (SIN/\milli\litre), to the \emph{actual} infectious virion (IV) concentration in the sample (\unit{10^6}{\IV/\milli\litre}), based on $10^5$ randomly simulated ED experiments, as a function of either (A) $\gamma$; (B) the ratio between the rate of loss of virus due to cell entry over that due to loss of infectivity, $\ssigma$; (C) $\burst=\rho\cdot\tau_I$; or (D) $n_I$. The experimentally measured SIN concentration reported for each simulated ED experiment corresponds to the midSIN-estimated highest likelihood \SIN/\milli\litre\ concentration for that ED outcome \cite{cresta21}. The black curves join the median (thick), 68\% (thin) and 95\% (dashed) credible interval (CI) of the vertically stacked histograms. The ED assay experiment parameters were $C_\text{input} = \unit{10^6}{IV/\milli\litre}$, $\mathcal{D}_1 = 10^{-2}$, dilution factor $= 10^{-0.6}$, $\Vinoc = \unit{0.1}{\milli\litre}$, $N_\text{cells}^\text{well} = \unit{10^5}{cells}$, using 11 dilutions and 8 replicate wells per dilution, as illustrated in Figure \ref{algo}. Infection parameters were varied about their base value ($\gamma=\unit{1}{cell/\IV}$, $(\beta N_\text{cells}/\Vol)/c=175$, $\burst=\unit{297}{\IV/cell}$, $n_I=60$, where $N_\text{cells}=N_\text{cells}^\text{well}$ and $\Vol=\Vinoc$) taken from \simon, with some adaptations for the new MM used herein (see Methods, Section \ref{mod}).%
}
\label{inf-params}
\end{figure}

In this example, the actual concentration of IVs in the sample is known, as are all the SM parameters. Normally, the SM parameters are unknown and meant to be estimated based on experimental SIN measurements whose relations to the actual number of infectious virions is therefore also unknown. Figure \ref{inf-params} explores how the SM parameters affect the experimentally measured \SIN/\milli\litre\ for a fixed, actual IV sample concentration of \unit{10^6}{\IV/\milli\litre}, based on ED assays randomly simulated as described above. As one decreases the probability ($\gamma$) that an IV's entry into a cell results in the cell's productive infection, an increasing number of IVs go uncounted by the ED assay, and the experimentally measured SIN increasingly underestimates the actual IV concentration in the sample (Figure \ref{inf-params}A). To a lesser degree, as $\ssigma$ decreases (Figure \ref{inf-params}B), there are fewer IV lost due to cell entry ($\beta N_\text{cells}/\Vol$) per IV losing infectivity ($c$), leading to fewer successful infections per infectious virion (smaller SIN/IV). In contrast, varying the mean and variance simultaneously (via $\burst = \rho\cdot\tau_I$) or only the variance (via $n_I$) of the distribution for the random total number of IV produced by each infected cell (Figure \ref{inf-params}C,D) has little to no effect on the number of measured infections per infectious virion (SIN/IV). This is in part due to the choice of SM parameters. More generally, if the mean number of IV produced by infected cells is too small, the infection of one cell might not lead to the infection of another. Similarly, lower $n_I$ values mean larger cell-to-cell variation in the total number of IV each cell produces, leading to a greater probability that an infected cell randomly produces little or no virus progeny, and thus fail to infect another cell \cite{quirouette23,yan20}. While $\burst$ and $n_I$ affect the likelihood of infection establishment only after a first successful cell infection, $\gamma$ and $\ssigma$ act to prevent this first infection from ever taking place, providing a stricter bottleneck, leading more easily to infection extinction.

\begin{figure}
\begin{center}
\includegraphics[width=0.8\linewidth]{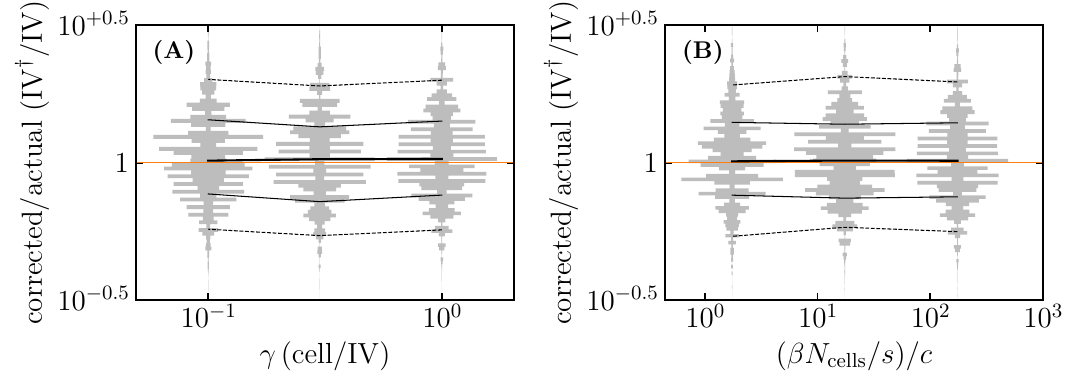}
\end{center}
\caption{%
\textbf{Recovering the actual infectious virus concentration by correcting the experimentally measured concentration.} %
Violin plot for the ratio of the \emph{corrected} experimentally measured virus concentration in IV$^\dagger$/\milli\litre, obtained by dividing the experimentally measured virus concentration in \SIN/\milli\litre\ by the establishment probability $\PVest$, over the \emph{actual} infectious virion concentration, based on simulated stochastic ED experiments, as a function of either (A) the probability of a successful cell infection post viral entry, $\gamma$; or (B) the ratio between the virus entry rate and the rate of loss of virus infectivity, $\ssigma$. Everything else is generated and represented visually as in Figure \ref{inf-params}.%
}
\label{inf-params-corrected}
\end{figure}

Perhaps unsurprisingly, the ratio of experimentally measured SIN per IV (Figure \ref{inf-params}A,B), corresponds exactly to the probability that a single infectious virion will lead to the successful establishment of the infection, where $\PVest = 1-(\PVext)$. In other words,
\begin{align} \label{conv}
\underbrace{C_\text{measured}}_{\SIN/\milli\litre} = \underbrace{C_\text{actual}}_{\IV/\milli\litre} \cdot \underbrace{\PVest}_{\SIN/\IV} \ . 
\end{align}
This is demonstrated in Figure \ref{inf-params-corrected} where the experimentally measured titres presented in Figure \ref{inf-params}A,B are divided by the establishment probability $\PVest$. This can be understood more intuitively when considering $\PVest$ for $n_I=1$, namely
\begin{align}
\PVest(n_I=1) = \left[\PVtoI\right] - \left[\frac{1}{\burst}\right] = \left[\frac{\gamma}{1 + c/(\beta N_\text{cells}/\Vol)}\right] - \left[\frac{1}{\rho\tau_I}\right]
\label{PVestRes}
\end{align}
Eqn.\ \eqref{PVestRes} shows why fewer infections per infectious virion (SIN/IV) are measured experimentally for low values of $\gamma$ or $\ssigma$. It also shows that the mean burst size, $\rho\tau_I$, affects $\PVest$ only when it is very small ($\rho\tau_I\sim 1$), and matters even less when $n_I>1$. As such, $\rho\tau_I$ is expected to play an important role only when virus production is suppressed, for example by the action of an antiviral drug \cite{quirouette23}. Similarly, small values of $n_I$ might not be biologically common or even realistic \cite{holder11,cbeau17,liao20}, such that $n_I$ likely rarely significantly affects $\PVest$. The average length of the eclipse phase ($\tau_E$), and the shape parameter of the Erlang-distributed eclipse phase duration ($n_E$), do not affect the extinction probability, and therefore have no impact on the ratio of measured infections per infectious virion.

\subsection{An alternative approach to parameter estimation}

Collectively, results thus far show that infection parameters affect the establishment probability and, therefore, the conversion of a virus sample's actual infectious virion (IV) concentration into its experimentally measured concentration of infection-causing units (SIN). This means that the parameters affect the experimentally measured viral titres that are in turn used to estimate these same parameters. The typical approach with ODE models circumvents this issue by expressing the MM variable corresponding to infectious viral titre in the same units as the experimental measurements ($V$ in PFU/\milli\litre\ or \tcid/\milli\litre\ or \SIN/\milli\litre). But as shown above, one experimentally measured infection-causing dose (e.g.,\ PFU, \tcid, SIN) could correspond to 10 or even 100 infectious virions, depending on $\PVext$. Consequently, important quantities such as the rate of mutation per infectious virion or the number of IV entry events required to cause the infection of one cell, cannot be determined, or only with poor accuracy, under some assumptions \cite{handel07,handel10,yan20,perelson12comb}.

Given these issues, we evaluate an alternative parameter estimation approach that takes into consideration the random nature of the virus infection process, the discrete nature of cells, virions, and infection events, as well as the random nature and limitations of experimental measurements of virus samples collected over the course of these infection. Compared to the standard approach, this alternative approach modifies three aspects of MM parameter estimation, namely
\begin{itemize}
\item using a stochastic model (SM) that describes the discrete number of infectious virions and cells in various states over the course of the in vitro virus infection, rather than an ODE model;
\item expressing infectious viral titre in the MM ($V$ in Eqn.\ \eqref{mfeq}) using biologically useful ($[V]=\IV$) rather than experimentally measurable ($[V]=\SIN$) units, and relating the MM variable in $\IV$ to the experimentally measured $\SIN$ via the establishment probability, $\PVest$, which explicitly takes consideration infection failure in the ED assay;
\item estimating MM parameters based on the likelihood of the experimentally observed ED assay outcome ($\LikeED$) for each measured sample (i.e.\ number of infected wells at each dilution), given the MM-predicted $\log_{10}(V)$ for a sample taken at that time, rather than using a likelihood that generically assumes normally distributed residuals ($\LikeNR$) between the experimentally measured and MM-predicted $\log_{10}(V)$.
\end{itemize}

In order to compare the standard parameter estimation approach (ODE with $[V]$ in $\SIN$ using $\LikeNR$) to this alternative approach (SM with $[V]$ in $\IV$ using $\LikeED$), we repeated the parameter estimation performed in \simon for the in vitro infection of A549 human lung carcinoma cells with a seasonal (A/New Caledonia/20-1999-like H1N1) influenza A virus (IAV) strain. Infections were performed in triplicate at two different virus inoculum concentrations (MOI of \unit{3}{PFU/cell} and \unit{0.01}{PFU/cell}) to produce a single-cycle (SC) and a multiple-cycle (MC) infection, with regular measurement of the concentration of infectious (via ED assay) and total (via qRT-PCR) viral titres \cite{simon16}. A Markov chain Monte Carlo (MCMC) approach was used to estimate the MM parameters' posterior distribution, given the experimental measurements. In the sections that follow, we describe in more details and individually evaluate the impact of the three changes.

\subsection{Likelihood of the experimentally measured infectious viral titres}

We compare two different expressions for the likelihood of the infectious viral titres measured via the ED assay, given the MM used and the set of parameters ($\pset$) to be estimated (see Methods for details). The standard one ($\LikeNR$) is based on the assumption that the residuals between the experimentally measured and MM-predicted $\log_{10}$ infectious titres follow a normal distribution,
\begin{align}
\LikeNR(\Cdata_r(t)|\sigma,\pset,\text{MM}) &=
\exp\left\{
- \frac{\left[\log_{10} \Cdata_r(t) -\log_{10} C^\text{model}(t|\pset) \right]^2}{2\sigma^2}
\right\} \ ,
\label{eqn:likenr}
\end{align}
where $\Cdata_r(t)$ is the experimentally measured concentration in $\SIN/\milli\litre$ estimated by midSIN \cite{cresta21} based on the ED assay performed on the sample taken at time $t$ from the $r^\text{th}$ infection replicate, $C^\text{model}(t|\pset)$ is the corresponding MM-predicted value, and $\sigma$ is the estimated standard deviation of the residuals (see Methods). The alternative ($\LikeED$) is the likelihood of the observed ED assay outcome, e.g.\ the number of positive wells at each dilution $(8,8,8,5,5,...)$, given the MM-predicted infectious titre for a sample taken at that time,
\begin{multline}
\LikeED(\vec{k}^\text{data}_r(t)|\vec{n}^\text{data}_r(t),\vec{\Dil}^\text{data}_r(t),\Vinoc,\pset,\text{MM}) = \\
\exp\left[-C^\text{model}(t|\pset) \cdot \Vinoc \cdot \sum_\col \Dil^\col_{r}(t) \left(n^\col_r(t)-k^\col_r(t)\right)\right] \\
\cdot \prod_{\col} \left(1-\exp\left[-C^\text{model}(t|\pset) \cdot \Vinoc \cdot \Dil^\col_r(t) \right] \right)^{k^\col_r(t)} \ ,
\label{eqn:likeed}
\end{multline}
where $\vec{k}^\text{data}_r(t)$ is a vector corresponding to the number of infected wells out of $\vec{n}^\text{data}_r(t)$ replicates in each column of the ED assay, given the total volume (sample plus diluant) placed in each well, $\Vinoc=\unit{0.05}{\milli\litre}$, and the sample's dilution factor in each column $\vec{\Dil}^\text{data}_r(t)$.  An example sample could be $\vec{k}^\text{data}_r(t)=(4,4,2,0,0,0,0,0)$, $\vec{n}^\text{data}_r(t) = (4,4,4,4,4,4,4,4)$, and $\vec{\Dil}^\text{data}_r(t) = (10^{-4},10^{-5},10^{-6},...,10^{-11})$, where $k^\col_r(t)=2$, $n^\col_r(t)=4$ and $\Dil^\col_r(t)=10^{-6}$ corresponds to the value of each of these quantities in the $3^\text{rd}$ (col=3) dilution column of the ED assay for one infection replicate ($r = 1$, 2, or 3), sampled at time $t$. The term $\exp[-C^\text{model}(t|\pset) \cdot \Vinoc]$ is the probability that a total volume $\Vinoc$ of the undiluted sample (if $\Dil^\col_r(t)=1$) placed in an ED well would fail to infect the well, given the MM-predicted infectious titre concentration, $C^\text{model}(t|\pset)$. Additional details about Eqn.\ \eqref{eqn:likeed} are provided in Methods.

Eqn.\ \eqref{eqn:likenr} for $\LikeNR$ reduces the outcome of an ED assay to a single value, namely the most likely $\SIN/\milli\litre$ concentration ($\Cdata_r(t)$), and assumes the same uncertainty for all measurements ($\sigma$). In contrast, Eqn.\ \eqref{eqn:likeed} for $\LikeED$ incorporates all the information provided by the ED assay, namely the number of infected and total wells at each dilution ($\vec{k}^\text{data}_r(t),\vec{n}^\text{data}_r(t)$), which includes information about the uncertainty of each measurement.

Another advantage of using $\LikeED$ is the ability to meaningfully include a likelihood for measurements that fall beyond the detection limit of the ED assay. Over the course of the triplicate MC infections, 6 of the 36 samples resulted in no infected well in the ED assay (0,0,0,...,0). The midSIN estimator provides an unnormalizable posterior distribution for such ED outcomes \cite{cresta21}, but it cannot provide a most likely $\SIN/\milli\litre$ concentration estimate, as is required by $\LikeNR$, because none exists. In handling these 6 measurements using $\LikeNR$, the residual is assumed to be zero ($\LikeNR=1$) when the MM-predicted $\SIN/\milli\litre$ concentration is less than or equal to the ED assay's lower limit of detection ($C^\text{model}(t|\pset) \le C_\text{LLoD}$), or is otherwise computed as the difference between the $\log_{10}$ MM-predicted titre and the assay's lower limit of detection, i.e.\ using $\Cdata_r(t)=C_\text{LLoD}$ in Eqn.\ \eqref{eqn:likenr}, where $C_\text{LLoD} = \unit{10^{1.494}}{\SIN/\milli\litre}$ (see Methods). In contrast, using the $\LikeED$, these 6 measurements can be treated in the same manner as any other measurement.

\begin{figure}
\includegraphics[width=1.00\linewidth]{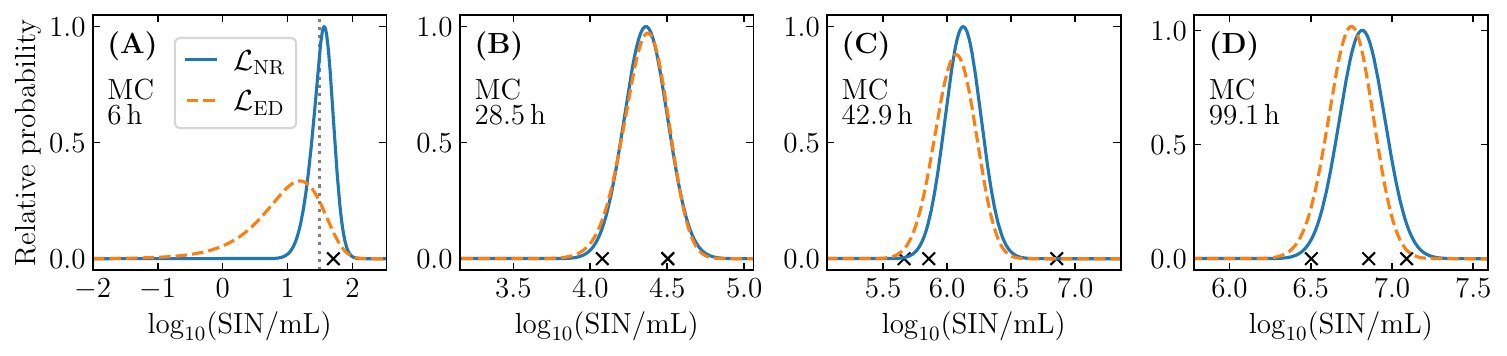}
\caption{%
\textbf{Visualizing the standard ($\LikeNR$) and alternative ($\LikeED$) likelihood functions.}
Relative probability ($y$ axis) of the three infectious titres measurements, given the $\log_{10}(\SIN/\milli\litre)$ predicted by the MM ($x$ axis), as computed according to either $\LikeNR$ (blue, solid) or $\LikeED$ (orange, dashed). The relative probability shown is the combined probability for all 3 experimental measurements taken at the indicated times, indicated as black $\times$ along the bottom of each graph. The vertical dashed line in (A) corresponds to the lower limit of detection of the ED assay ($C_\text{LLoD}$, Table \ref{tab:fixdpars}). A comparison of $\LikeNR$ vs $\LikeED$ for all measurements taken over the SC and MC infections is provided in Methods (Figures \ref{likeliMC} and \ref{likeliSC}).
}
\label{fig:likedata}
\end{figure}

Figure \ref{fig:likedata} shows the probability ($y$-axis), computed using either $\LikeNR$ or $\LikeED$, that the $\log_{10}$ SIN concentration measured at a particular time $t$ takes on a certain value ($x$ axis), given the three experimental measurements from replicate MC infections taken at different times. Figure \ref{fig:likedata}A shows the impact of having 2 out of 3 replicate measurements below the lower limit of detection. The $\LikeNR$ is a poor match for the correct $\LikeED$. Figure \ref{fig:likedata}B shows that $\LikeNR$ can sometimes closely match $\LikeED$. Figure \ref{fig:likedata}C,D suggest that the standard deviation, assumed in $\LikeNR$ to be the same for all MC infectious titre measurements ($\sigma_\mathrm{MC}$, Table \ref{tab:fixdpars}), is at times smaller (Figure \ref{fig:likedata}C) or larger (Figure \ref{fig:likedata}D) than the correct, $\LikeED$-computed standard deviation. Of the 24 experimental samples taken over the course of the triplicate SC and MC infections (see Methods), $\LikeNR$ always overestimated the most likely $\log_{10}(\SIN/\milli\litre)$ (e.g.\ by 2.3-fold in A, 11\% in C, 15\% in D), with the single exception of the measurement shown in Figure \ref{fig:likedata}B, where it underestimates it by 2.6\%.

\begin{figure}
\begin{center}
\includegraphics[width=1.00\linewidth]{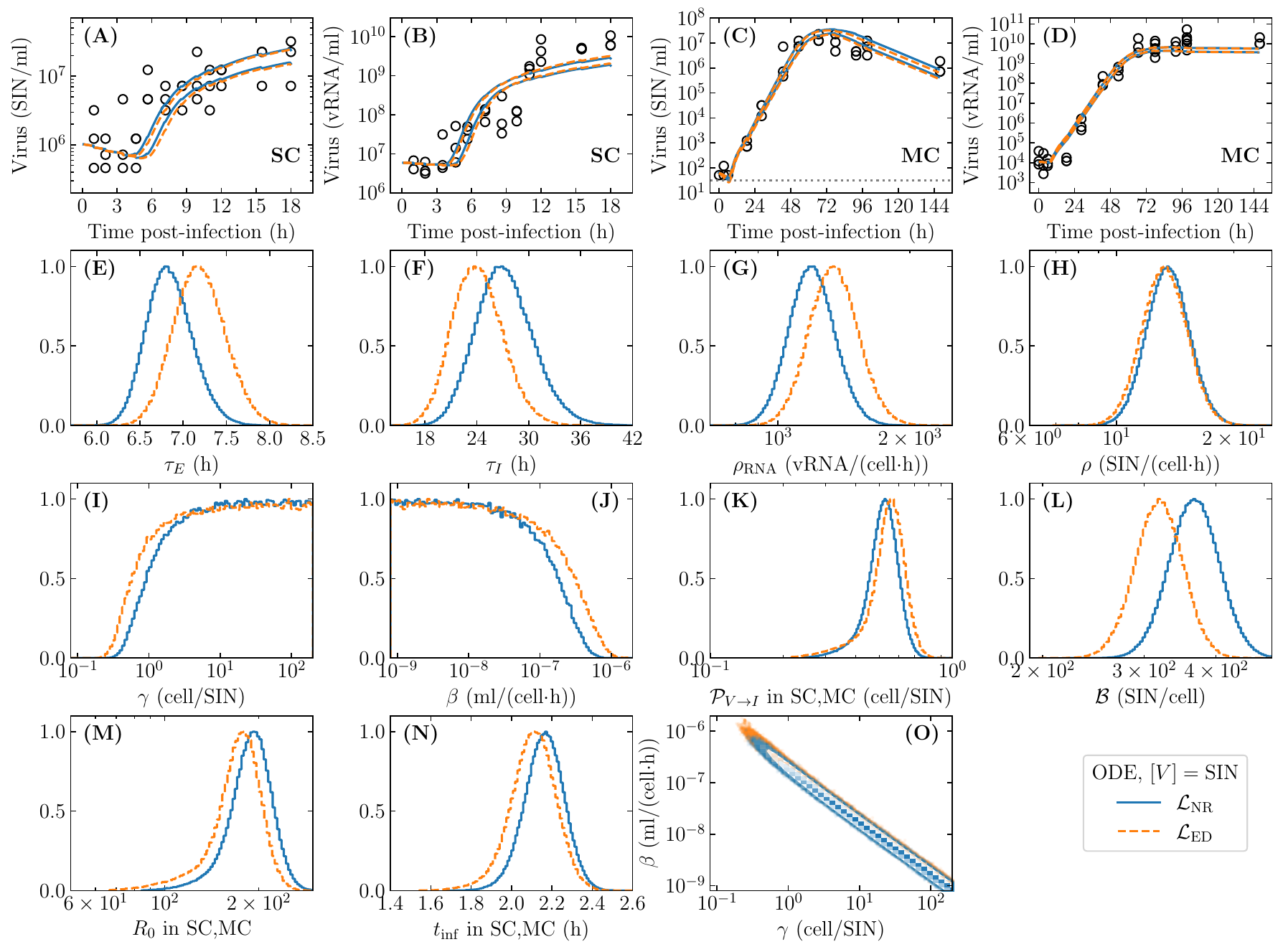}
\end{center}
\caption{%
\textbf{Impact of the likelihood function on parameter estimation.} %
(A--D) Triplicate infectious (A,C) or total (B,D) viral titre measurements (circles) over the course of triplicate SC (A,B) or MC (C,D) in vitro influenza A virus infections, along with the 95\% CI for the MM-predicted viral time courses, based on using either the $\LikeNR$ (solid blue line) or $\LikeED$ (dashed orange line) likelihood. %
(E--J) Marginalized posterior distributions (MPDs) of the MCMC-estimated parameters, where $\tau_E$ and $\tau_I$ are the mean duration of the eclipse and infectious phases (h), $\rho_\text{RNA}$ and $\rho$ are the total and infectious viral production rates (vRNA/h and $\SIN$/h), $\gamma$ is the number of cells infected per $\SIN$ lost due to cell entry (cell/$\SIN$) and $\beta$ is the rate of $\SIN$ loss due to cell entry ($\milli\litre/[\text{cell}\cdot\hour]$). %
(K--N) MPDs of biologically relevant quantities derived from the MCMC-estimated parameters, where $\burst=\rho\tau_I$ is the mean number of $\SIN$ produced per cell, $\PVtoI = \gamma/[1+(c/\beta\Ncells/\Vol)]$ is the mean number of cells infected per $\SIN$ produced (cell/$\SIN$), $\tinf$ is the infecting time (see Methods), and $R_0$ is the basic reproductive number. %
(O) Paired MPD for correlated parameters $\gamma$ and $\beta$.
}
\label{fig:varylike}
\end{figure}

Figure \ref{fig:varylike}A--D shows the ODE-predicted infectious and total viral titres over the course of the SC and MC infections, and Figure \ref{fig:varylike}E--J the corresponding marginalized posterior distributions (MPDs) for the 6 parameters estimated based on either $\LikeNR$ or $\LikeED$. All parameters have similar but not identical estimated values (Figure \ref{fig:varylike}E--J). The $\LikeED$ estimates a longer eclipse phase ($\tau_E$, Figure \ref{fig:varylike}E), shorter infectious phase ($\tau_I$, Figure \ref{fig:varylike}F), and higher total viral production rate ($\rho_\text{RNA}$, Figure \ref{fig:varylike}G) than that obtained using the standard $\LikeNR$. Notably, the effect of the longer eclipse phase is visible as a delay (right-shift) of the ODE-predicted viral curves in the SC infection (Figure \ref{fig:varylike}A,B), and the shorter infectious phase appears as a shorter viral plateau and a more rapid transition to viral decay in the ODE-predicted MC infection time course (Figure \ref{fig:varylike}C). These differences are not statistically significant (Tables \ref{tab4pars} and \ref{tab4pvals}) and are unlikely to be biologically significant in comparison to typical inter-experimental variability \cite{paradis15}.

\begin{sidewaystable}
\caption{\textbf{Parameters estimated over four variations$^a$}}
\label{tab4pars}
\begin{center}
\begin{tabular}{lllll}
\hline
Parameters & MFM,$\SIN$,$\LikeNR$ & MFM,$\SIN$,$\LikeED$ & MFM,$\IV$,$\LikeED$ & SM,$\IV$,$\LikeED$ \\
\hline
$\tau_E$ (h) & $6.8\,[6.3,7.4]\,(6.8)$ & $7.2\,[6.6,7.8]\,(7.1)$ & $7.2\,[6.6,7.8]\,(7.2)$ & $7.1\,[6.6,7.7]\,(7)$ \\
$\tau_I$ (h) & $27\,[21,34]\,(28)$ & $24\,[19,30]\,(25)$ & $24\,[19,30]\,(24)$ & $24\,[19,30]\,(25)$ \\
$\rho_\mathrm{RNA}$ (vRNA/(cell$\cdot$h)) & $10^{3.1\,[3,3.2]\,(3.1)}$ & $10^{3.1\,[3,3.2]\,(3.1)}$ & $10^{3.1\,[3,3.2]\,(3.1)}$ & $10^{3.1\,[3,3.2]\,(3.1)}$ \\
$\rho$ ($[V]$/(cell$\cdot$h)) & $10^{1.1\,[1,1.2]\,(1.1)}$ & $10^{1.1\,[1,1.2]\,(1.1)}$ & $10^{1.7\,[1.2,1.9]\,(1.9)}$ & $10^{1.7\,[1.2,1.9]\,(1.7)}$ \\
$\gamma$ (cell/$[V]$) & $10^{2.9\,[0.026,3.2]\,(-0.2)}$ & $10^{2.4\,[-0.15,3.2]\,(-0.24)}$ & $10^{-0.6\,[-0.67,-0.016]\,(-0.68)}$ & $10^{-0.64\,[-0.69,-0.028]\,(-0.5)}$ \\
$\beta$ (ml/(cell$\cdot$h)) & $10^{-9.7\,[-10,-6.8]\,(-6.5)}$ & $10^{-9.1\,[-10,-6.6]\,(-6.4)}$ & $10^{-6.8\,[-6.9,-6.7]\,(-6.8)}$ & $10^{-6.8\,[-6.9,-6.6]\,(-6.7)}$ \\
\hline
\multicolumn{5}{l}{Derived quantities} \\
\hline
$\mathcal{B}$ ($[V]$/cell) & $10^{2.6\,[2.5,2.6]\,(2.6)}$ & $10^{2.5\,[2.4,2.6]\,(2.5)}$ & $10^{3.1\,[2.6,3.3]\,(3.3)}$ & $10^{3.2\,[2.6,3.3]\,(3.1)}$ \\
$\mathcal{P}_{V\rightarrow I}$ in SC,MC (cell/$[V]$) & $10^{-0.28\,[-0.44,-0.16]\,(-0.5)}$ & $10^{-0.25\,[-0.47,-0.13]\,(-0.5)}$ & $10^{-1.1\,[-1.1,-0.46]\,(-1.1)}$ & $10^{-1\,[-1.1,-0.45]\,(-0.89)}$ \\
$\mathcal{P}_{V\rightarrow I}$ in ED (cell/$[V]$) & $10^{0.73\,[-0.026,0.86]\,(-0.24)}$ & $10^{0.76\,[-0.17,0.89]\,(-0.28)}$ & $10^{-0.67\,[-0.76,-0.11]\,(-0.75)}$ & $10^{-0.71\,[-0.76,-0.11]\,(-0.55)}$ \\
$R_0$ in SC,MC & $10^{2.3\,[2.1,2.4]\,(2.1)}$ & $10^{2.3\,[2,2.4]\,(2)}$ & $10^{2.1\,[2,2.2]\,(2.1)}$ & $10^{2.1\,[2,2.2]\,(2.2)}$ \\
$R_0$ in ED & $10^{3.3\,[2.5,3.4]\,(2.3)}$ & $10^{3.3\,[2.3,3.4]\,(2.2)}$ & $10^{2.5\,[2.4,2.6]\,(2.5)}$ & $10^{2.5\,[2.4,2.6]\,(2.5)}$ \\
$t_\mathrm{inf}$ in SC,MC (h) & $2.2\,[2,2.4]\,(2)$ & $2.1\,[1.9,2.3]\,(1.9)$ & $2\,[1.8,2.2]\,(2)$ & $1.8\,[1.6,1.9]\,(1.7)$ \\
$t_\mathrm{inf}$ in ED (h) & $0.67\,[0.6,0.72]\,(0.62)$ & $0.65\,[0.58,0.72]\,(0.6)$ & $0.62\,[0.57,0.68]\,(0.62)$ & $0.54\,[0.49,0.6]\,(0.52)$ \\
$\mathcal{P}_{V\rightarrow \mathrm{Est.}}$ in SC,MC & --- & --- & $10^{-1.1\,[-1.1,-0.46]\,(-1.1)}$ & $10^{-1\,[-1.1,-0.45]\,(-0.89)}$ \\
$\mathcal{P}_{V\rightarrow \mathrm{Est.}}$ in ED & --- & --- & $10^{-0.67\,[-0.76,-0.11]\,(-0.75)}$ & $10^{-0.71\,[-0.76,-0.11]\,(-0.55)}$ \\
\hline

\end{tabular}
\begin{minipage}{0.95\linewidth}
$^a$ Estimates are provided as: Mode [95\% CI] (MAP) where the mode and 95\% CI correspond to the MPD for each parameter, marginalized over all others, and the MAP corresponds to the parameter set $\pset$ with the highest likelihood in the joint, multi-dimensional posterior $\mathcal{P}_\text{post}(\pset|\text{data})$. Two values are provided for the quantities $\PVtoI$, $R_0$, $\tinf$, $\PVest$ which reflect the different cell density in the SC and MC infection experiments ($\Ncells/\Vol$) and those in the ED assay ($N_\text{cells,ED}/\Vinoc$).
\end{minipage}
\end{center}
\end{sidewaystable}

\begin{table}
\caption{\textbf{$p$-values for pairwise comparison of variations$^a$}}
\label{tab4pvals}
\begin{center}
\begin{tabular}{lcccc}
\hline
Parameters & $\LikeNR$ vs.\ $\LikeED$ & $\SIN$ vs.\ $\IV$ & MFM vs.\ SM & old vs.\ new$^b$ \\
\hline
$\tau_E$ (h) & 0.19 & 0.5 & 0.43 & 0.25 \\
$\tau_I$ (h) & 0.24 & 0.5 & 0.47 & 0.26 \\
$\rho_\mathrm{RNA}$ (vRNA/(cell$\cdot$h)) & 0.26 & 0.49 & 0.47 & 0.29 \\
$\rho$ ($[V]$/(cell$\cdot$h)) & 0.46 & --- & 0.49 & --- \\
$\gamma$ (cell/$[V]$) & 0.48 & --- & 0.48 & --- \\
$\beta$ (ml/(cell$\cdot$h)) & 0.47 & --- & 0.39 & --- \\
\hline
\multicolumn{5}{l}{Derived quantities} \\
\hline
$\mathcal{B}$ ($[V]$/cell) & 0.15 & --- & 0.49 & --- \\
$\mathcal{P}_{V\rightarrow I}$ in SC,MC (cell/$[V]$) & 0.41 & --- & 0.49 & --- \\
$\mathcal{P}_{V\rightarrow I}$ in ED (cell/$[V]$) & 0.47 & --- & 0.49 & --- \\
$R_0$ in SC,MC & 0.34 & 0.12 & 0.41 & 0.058 \\
$R_0$ in ED & 0.4 & 0.09 & 0.49 & \textbf{0.039} \\
$t_\mathrm{inf}$ in SC,MC (h) & 0.36 & 0.29 & \textbf{0.022} & \textbf{0.0023} \\
$t_\mathrm{inf}$ in ED (h) & 0.36 & 0.29 & \textbf{0.022} & \textbf{0.0023} \\
$\mathcal{P}_{V\rightarrow \mathrm{Est.}}$ in SC,MC & --- & --- & 0.49 & --- \\
$\mathcal{P}_{V\rightarrow \mathrm{Est.}}$ in ED & --- & --- & 0.49 & --- \\
\hline

\end{tabular}
\begin{minipage}{0.95\linewidth}
$^a$ The $p$-value corresponds to the fraction of times the statement `$A > B$' or `$A < B$' is true (not both, i.e., one-tailed) for 1,000,000 pairwise comparisons of individual parameter values drawn at random from $A$ and $B$ with replacement, where $A$ and $B$ are the sets of MCMC-accepted parameters obtained using similar procedures that differed only in the aspect indicated in the column header. \\
$^b$ Old refers to ($\LikeNR$, $[V]=\SIN$, ODE) and new to ($\LikeED$, $[V]=\IV$, SM).
\end{minipage}
\end{center}
\end{table}

Figure \ref{fig:varylike}K--N shows 4 biologically relevant quantities computed from the ODE-estimated parameters. Figure \ref{fig:varylike}K corresponds to $\PVtoI$ (see Figure \ref{figMM}), the probability that a cell becomes productively infected by one IV when $[V]$ is expressed in units of $\IV$. Since here $[V]=\SIN$, and one $\SIN$ (observed infection) can represent several $\IV$ (infectious virions), it is possible for one SIN to infect more than one cell. As such, when $[V]=\SIN$, $\PVtoI$ can be larger than one (no longer a probability), and corresponds to the mean number of cells infected per SIN produced ($R_0/\burst$, with units of cell/SIN). The product of $\burst$ (mean total SIN progeny per infected cell) and $\PVtoI$ (cell infected per SIN produced) corresponds to $R_0$, the basic reproductive number. Use of $\LikeED$ resulted in a larger $\PVtoI$, smaller $\burst$, and a smaller $R_0$ (Figure \ref{fig:varylike}K,L,M, respectively), but these differences are not significant (Table \ref{tab4pars},\ref{tab4pvals}). The infecting time ($\tinf$), first introduced in Holder et al.\ \cite{holder11}, represents the time for a newly infectious cell to cause the infection of its first cell, given an otherwise fully susceptible cell population, and neglecting virus loss of infectivity. Herein, the definition for $t_\text{inf}$ neglects both virus loss of infectivity and loss to cell entry, and depends on $\rho\cdot\gamma\beta\Ncells/\Vol$ in the SC and MC infections (see Methods). It is estimated to be slightly shorter using $\LikeED$ than $\LikeNR$ (Figure \ref{fig:varylike}N), but not significantly so (Table \ref{tab4pars},\ref{tab4pvals}).

Parameter estimation with either likelihood functions results in a correlation (Figure \ref{fig:varylike}O) between the rate of $\SIN$ loss due to cell entry ($\beta$, Figure \ref{fig:varylike}J), and the number of cells infected per $\SIN$ cell entry ($\gamma$, Figure \ref{fig:varylike}I). Even though $\gamma$ and $\beta$ are poorly constrained by the data (a lower bound of $\beta\ge\unit{10^{-10}}{\milli\litre/(cell\cdot\hour)}$ had to be imposed), the derived quantities which depend on these parameters ($\PVtoI$, $R_0$, $\tinf$) are comparatively well-constrained. The ($\beta$,$\gamma$) correlation, and the reason for the need to impose a lower bound on $\beta$, are discussed in the next section. Paired MPDs for all parameter pairs using the $\LikeNR$ or $\LikeED$ are provided in Methods.

\subsection{Units used to express the infectious viral titre in the MM}
\label{units}

ODE models typically express the infectious viral titre, variable $V$ in Eqn.\ \eqref{mfeq} (see Methods), using the same units as the experimentally measured infectious titre (e.g., PFU/\milli\litre, \tcid/\milli\litre, \SIN/\milli\litre) against which $V$ is compared to estimate the MM parameters. In contrast, SMs track individual virions and cells, expressing these variables as whole numbers, and as such $V$ in SMs corresponds to the number of infectious virions (\IV). This requires converting the units of $V$ (hereafter $[V]$) from $\IV$ to $\SIN$ to be compared against the experimentally measured infectious titre in $\SIN$, which can be challenging \cite{handel07}. Fortunately, Eqn.\ \eqref{conv} shows how this conversion can be performed using the infection establishment probability, $\PVest$.

To explore the impact of representing $V$ in units of infectious virions ($[V]=\IV$) rather than SIN ($[V]=\SIN$), we repeat the parameter estimation with the ODE model using $\LikeED$, this time using $[V]=\IV$. In the previous section, where $[V]=\SIN$, the ODE-predicted SIN concentration at time $t$ in the $\LikeED$ likelihood (Eqn.\ \eqref{eqn:likeed}) corresponds to $C^\text{model}(t) = V(t)/\Vol$, where $\Vol=\unit{10}{\milli\litre}$ is the total volume of supernatant in the SC and MC in vitro infections. When instead $[V]=\IV$, $C^\text{model}(t) = [V(t)/\Vol\cdot\PVest(\pset)]$, where $\PVest(\pset)$ is the infection establishment probability in the ED assay where the sample is measured, not that in the SC and MC infections from which it was drawn. If the SC and MC infections are performed under the same conditions as the ED assays (same cells, temperature, sample diluent, etc.), they should share MM parameters (have the same $\pset$), except often for the number of cells and the volume of supernatant, which tend to differ between infection and measurement assays, but are known rather than estimated (see Methods, Table \ref{tab:fixdpars}). Unfortunately, the SC and MC infections considered here were performed in A549 cells, whereas the ED assays were performed using MDCK cells. For simplicity, we assume the SC and MC infection experiments and the ED assay differ only in the number of cells ($\Ncells$ vs $N_\text{cells,ED}$) and supernatant volume ($\Vol$ vs $\Vinoc$), as detailed in Methods. This assumption is necessary because the ED assay alone does not provides sufficient information to estimate its MM parameters. It is nonetheless incorrect, and its impact will be explored in the Discussion.

\begin{figure}
\begin{center}
\includegraphics[width=1.00\linewidth]{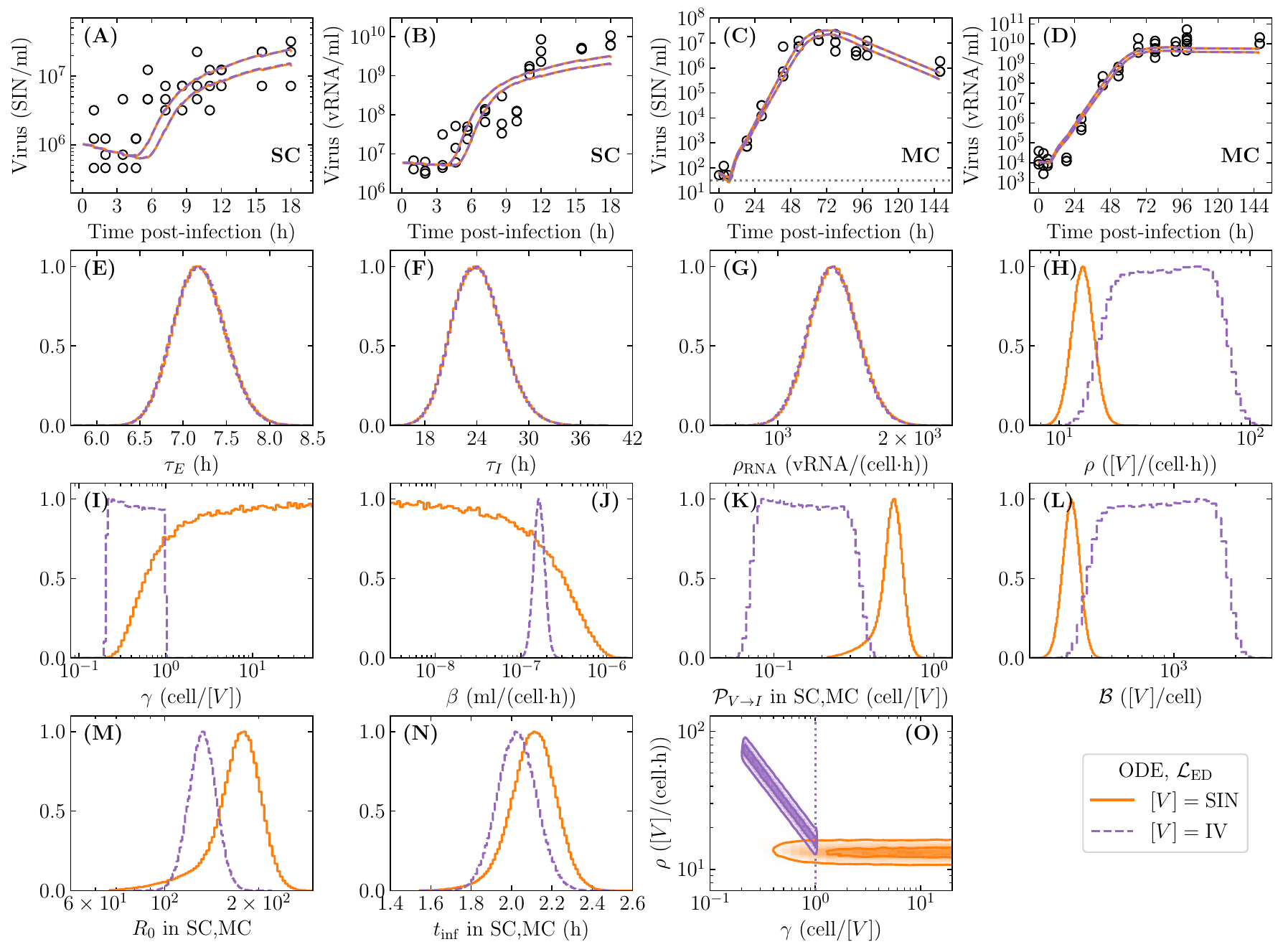}
\end{center}
\caption{%
\textbf{Impact of the infectious viral titre ($V$) units on parameter estimation.} %
MM-predicted viral time courses and MPDs for the ODE model, using the $\LikeED$, when $[V]=\SIN$ (solid orange) or $[V]=\IV$ (dashed purple).
Panels (A--N) are as described in Figure \ref{fig:varylike}, although a number of parameters ($\rho$, $\gamma$, $\beta$, $\PVtoI$, $\burst$, and $\PVest$) have a different biological meaning depending on whether $[V]=\SIN$ or $\IV$, as explained in the text.
(O) Paired MPD for $\gamma$ and $\rho$, which shows a correlation only when $[V]=\IV$, which physically requires that $\gamma\le\unit{1}{infected\ cell}$ per $\IV$ entry into cell (vertical dashed line).
}
\label{fig:varyunits}
\end{figure}

Figure \ref{fig:varyunits} shows the impact of expressing $V$ in the ODE model in units of $\SIN$, as is typically done, rather than in units of $\IV$. The change of units has no effect on the ODE-predicted total and infectious viral titre time courses (Figure \ref{fig:varyunits}A--D), nor on the estimated value of parameters $\tau_E$, $\tau_I$, and $\rho_\text{RNA}$ (Figure \ref{fig:varyunits}E--G, Table \ref{tab4pars},\ref{tab4pvals}). It affects estimates of the production rate of $V$ ($\rho$ in $[V]/(\text{cell}\cdot\hour)$), the rate of loss of $V$ due to cell entry ($\beta$ in $\milli\litre\cdot(\text{cell}\cdot\hour)^{-1}$), and the number of cells infected per $V$ entry ($\gamma$ in $\text{cell}/[V]$). This is because these parameters have different units, or at least a different biological meaning, depending on whether $[V]=\SIN$ or $\IV$, and thus cannot be compared.

\begin{figure}
\begin{center}
\includegraphics[width=\linewidth]{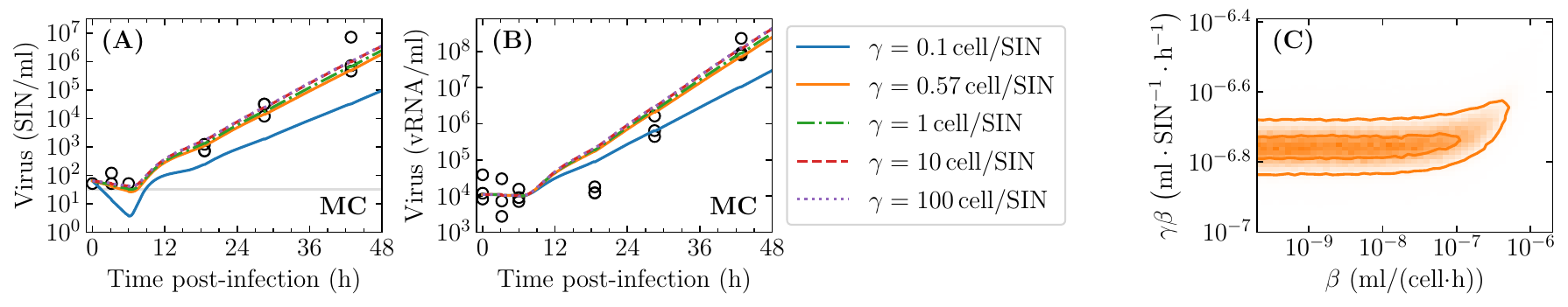}
\end{center}
\caption{%
\textbf{Parameter degeneracy between $\gamma$ and $\beta$ when $V$ is in units of $\SIN$.} %
ODE-predicted (A) $\SIN$ and (B) total viral concentration time course for the MC infection over the first $\unit{48}{hpi}$ using the maximum a posteriori (MAP) parameter set (Table \ref{tab4pars}, ODE,$\SIN$,$\LikeED$) where $\gamma$ and $\beta$ are modified so as to keep their product ($\gamma\beta$) constant while $\gamma$ is varied.
(C) Paired MPD for the product $\gamma\beta$ as a function of $\beta$, based on an imposed artificial lower bound requiring $\beta\ge\unit{10^{-10}}{\milli\litre\cdot(\mathrm{cell}\cdot\hour)^{-1}}$.
}
\label{fig:gb_correls}
\end{figure}

When $[V]=\SIN$, the $\SIN$ production rate ($\rho$, Figure \ref{fig:varyunits}H) is well constrained by the experimental data, but $\beta$ and $\gamma$ are correlated (Figure \ref{fig:varylike}O). As $\gamma$ is decreased and $\beta$ is increased so as to keep their product ($\gamma\beta$) fixed, the rate of cell infection per $\SIN$ cell entry ($\gamma\beta TV/\Vol$) is unchanged, but the rate of $\SIN$ loss to cell entry ($\beta TV/\Vol$) increases. This results in a slower $\SIN$ and total viral titer growth rate and an increasingly more pronounced $\SIN$ titre loss at the start of the MC infection (Figure \ref{fig:gb_correls}A,B), when the MOI is low and the number of available target cells absorbing $\SIN$ is large ($T\sim\Ncells$). Experimentally measured viral titres at these early time points therefore provide an upper bound on $\beta$, and a corresponding lower bound on $\gamma$. In the other direction, as $\beta$ decreases, the rate of $\SIN$ loss due to cell entry ($\beta TV/\Vol$) becomes negligible, and the rate of cell infection ($\gamma\beta TV/\Vol$) depends only on the product $\gamma\beta$: their individual value no longer matters (Figure \ref{fig:gb_correls}C). Therefore, when $[V]=\SIN$, it is necessary to impose a lower bound on $\beta$ (upper bound on $\gamma$) to allow for their MPDs to be finite and therefore normalizable, albeit incorrect.

When the establishment probability of an infection initiated with a single infectious virion is less than one ($\PVest<\unit{1}{\SIN/\IV}$), one infectious dose (\unit{1}{\SIN}) represents more than one infectious virion (\unit{1}{\IV}). This makes it possible for one SIN to infect more than one cell ($\gamma > \unit{1}{cell/SIN}$), since one SIN represents more than one IV. Previous work analyzing the same data presented herein with a similar ODE model estimated a 95\% CI for $\gamma$ (parameter $1/n$ therein) of \unit{[1.2,121]}{cell/\tcid} \cite{yan20}. Other work based on influenza A virus infections in mice estimated values in the range \unit{[7,10^3]}{cell/PFU} based on multiple data sets \cite{handel10}. Both works assumed that $\gamma\ge\unit{1}{cell}$ per infectious titre unit (e.g., PFU, \tcid), i.e.\ that at least one cell or more should be infected per infectious dose lost to cell entry. But the entry of one $\SIN$ into a cell could also occasionally fail to cause its infection such that $\gamma$ could also plausibly be slightly less than $\unit{1}{cell/\SIN}$. Here, it is the data which imposes the lower bound on $\gamma$ (upper bound on $\beta$), while the artificially imposed lower bound on $\beta$ ($\geq \unit{10^{-10}}{ml/(cell \cdot h)}$) was chosen to be low enough that a further decrease in $\beta$ does not meaningfully change the posterior distribution (flat MPD once $\beta\lessapprox\unit{10^{-8}}{ml/(cell \cdot h)}$, Figures \ref{fig:varylike}J, \ref{fig:gb_correls}C).

\begin{figure}
\begin{center}
\includegraphics[width=0.75\linewidth]{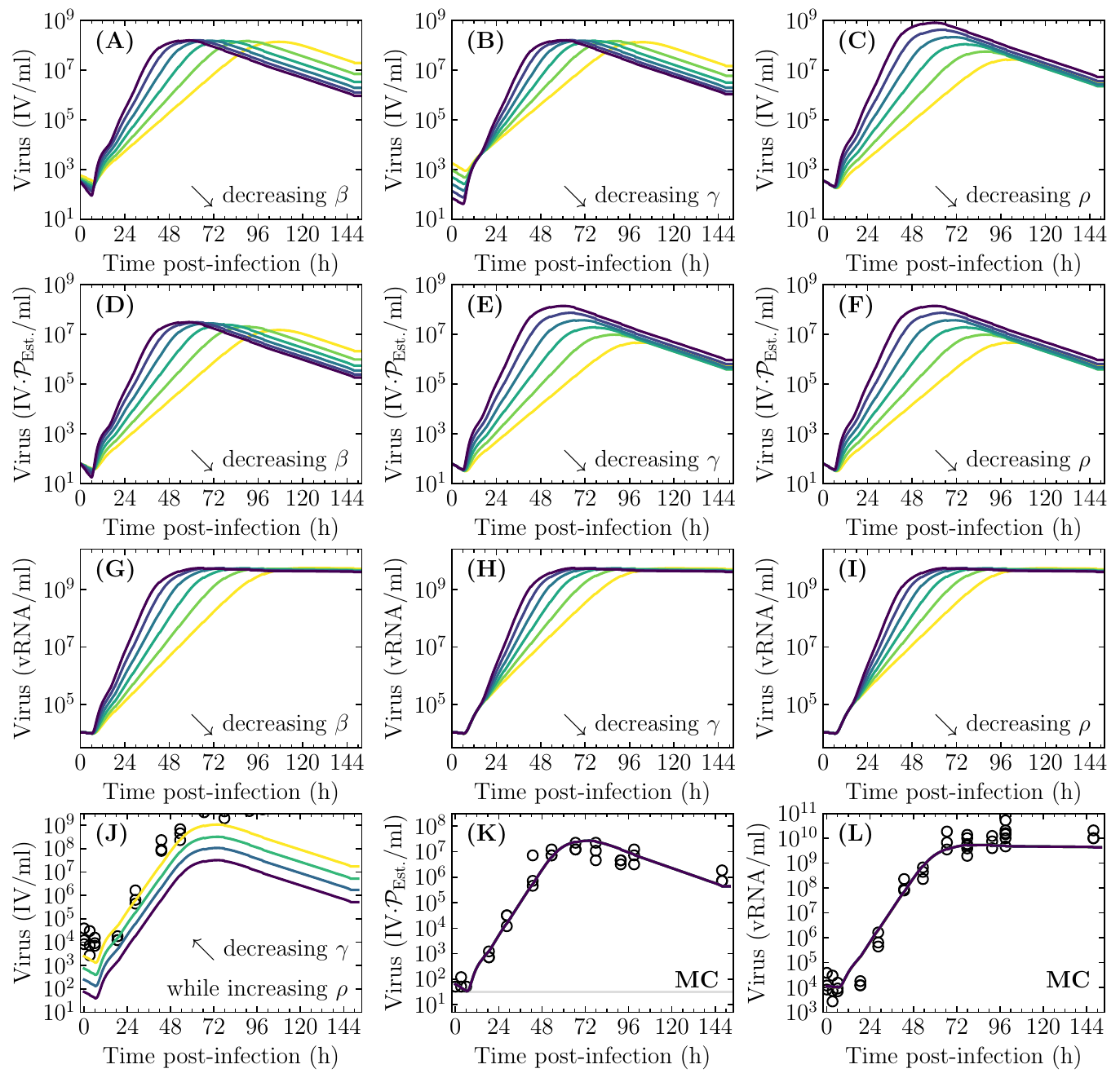}
\end{center}
\caption{%
\textbf{Parameter degeneracy between $\gamma$ and $\rho$ when $V$ is in units of $\IV$.} %
ODE-predicted (A,B,C) $\IV$, (D,E,F) $\SIN$ and (G,H,I) total viral titre concentration over the course of the MC infection using the MAP parameter set (Table \ref{tab4pars}, ODE,$\IV$,$\LikeED$), and varying (A,D,G) $\beta$, (B,E,H) $\gamma$ or (C,F,I) $\rho$ by factors log-uniformly spaced over $[10^{-0.7},10^{0.7}]$ about their MAP value. 
ODE-predicted (J) $\IV$, (K) $\SIN$, and (L) total viral titre concentration when varying $\gamma=\unit{[0.03,0.1,0.3,1.0]}{cell/\IV}$ and $\rho$ simultaneously so as to keep their product ($\gamma\rho$) constant. The (K) infectious ($\SIN/\milli\litre$) and (J,L) total (vRNA/$\milli\litre$) viral titre (open circles) experimentally measured over the course of the triplicate MC infections are also shown for comparison.
In (D,E,F,K), the concentration in $\SIN$ is obtained by converting the ODE-predicted time courses in (A,B,C,J) from $\IV$ to $\SIN$ by multiplying $V$ in $\IV$ by the establishment probability ($\PVest(\pset)$, Eqn.\ \eqref{conv}).
}
\label{fig:gp_correls}
\end{figure}

When instead $[V]=\IV$, the rate of infectious virions loss to cell entry is well constrained ($\beta$, Figure \ref{fig:varyunits}J), but $\rho$ and $\gamma$ are correlated (Figure \ref{fig:varyunits}H,I,O). Figure \ref{fig:gp_correls} shows the effect of individually varying $\beta$, $\rho$, or $\gamma$ on the ODE-predicted MC infection time course in $\IV$ (Figure \ref{fig:gp_correls}A--C), their conversion to $\SIN$ (Figure \ref{fig:gp_correls}D--F), and the total virus concentration (Figure \ref{fig:gp_correls}G--I). While ($\gamma,\beta$) similarly affect the growth rate of the ODE-predicted $V$ in $\IV$ (Figure \ref{fig:gp_correls}A,B), the $\IV$ production rate ($\rho$) affects both the growth rate and the peak titre (Figure \ref{fig:gp_correls}C). This is the same ($\gamma,\beta$) degeneracy reported when $V$ is directly expressed in $\SIN$ (Figure \ref{fig:varylike}O and \ref{fig:gb_correls}C). The shift from a ($\gamma,\beta$) to a ($\gamma,\rho$) correlation (Figures \ref{fig:varylike}O to \ref{fig:varyunits}O) arises when the ODE-predicted $V$ is converted from infectious virions to measured infections ($\IV$ to $\SIN$) to be compared against the experimentally measured $\SIN$ titres. This is accomplished by multiplying the number of infectious virions ($\IV$) by the establishment probability ($\PVest$) which varies as $\beta$, $\rho$, or $\gamma$ are varied. When $\burst=\rho\tau_I\gg1$, as is the case here (Figure \ref{fig:varyunits}L), the productive infection of one cell essentially guarantees the establishment of the infection, such that the probability that the infection establishes depends only on the probability that one IV productively infects a cell ($\PVest \approx \PVtoI$, Figure \ref{fig:varyunits}K, Table \ref{tab4pars}). Rewriting $\PVtoI$ as $\gamma(\beta\Ncells/\Vol)/[(\beta\Ncells/\Vol)+c]$ shows why varying $\beta$, appearing in both the numerator and denominator, has a marginal impact on $\PVest$ and on the conversion of $\IV$ to $\SIN$ (e.g., no impact on peak titre, Figure \ref{fig:gp_correls}A vs.\ D). In contrast, reducing $\gamma$ by some factor, reduces $\PVest$ and the conversion of the $V$ curve to $\SIN$ by approximately the same factor (e.g., same fold decrease in peak titre, Figure \ref{fig:gp_correls}B,E). Thus, after the ODE-predicted curve in $\IV$ is converted to $\SIN$, $\gamma$ has the same effect as $\rho$ (Figure \ref{fig:gp_correls}E,F), altering both the growth rate and peak titre, resulting in a ($\gamma$,$\rho$) correlation (Figure \ref{fig:varyunits}O). Since viral titre peak, after conversion to $\SIN$, is proportional to both $\gamma$ and $\rho$, a 3-fold decrease in $\gamma$ with a corresponding 3-fold increase in $\rho$, wherein $\gamma\rho$ remains constant, shifts the $\IV$ curve upwards (higher $\IV$ peak, Figure \ref{fig:gp_correls}J) but leaves the converted $\SIN$ curve unchanged (Figure \ref{fig:gp_correls}K).

Importantly, when $[V]=\IV$, it becomes possible to impose appropriate physical constraints on $\gamma$, the number of cells infected per $\IV$ lost to cell entry. At most $\gamma\le\unit{1}{cell/\IV}$ since it is impossible for a cell to become infected without the entry of at least one $\IV$ into the cell, setting a corresponding lower bound on the IV production rate by infected cells, $\rho$. In the other direction, as $\gamma$ and thus $\PVest$ decrease, $\rho$ and the ODE-predicted $\IV$ curve get increasingly high (Figure \ref{fig:gp_correls}J) so that the product of $V$ in $\IV$ and $\PVest$ continue to match the experimentally measured titres in $\SIN$ (Figure \ref{fig:gp_correls}K). But $\gamma$ cannot decrease (and the $\IV$ curve shift upward) indefinitely because, physically, the ODE-predicted number of $\IV$ cannot exceed the total number of experimentally observed virions (infectious and not), i.e.\ the number of viral RNA copies measured via qRT-PCR (Figure \ref{fig:gp_correls}K,L). This lower bound on $\gamma$ (upper bound on $\rho$) is imposed by rejecting (assigning a likelihood of zero to) any parameter set ($\pset$) that results in an ODE-predicted initial $\IV$ inoculum ($V$) exceeding the experimentally measured number of viral RNA at that time ($V\rna$), as part of the MCMC process to estimate the MM parameters when $[V]=\IV$. We furthermore impose that $\rho \leq \rho_\text{RNA}$, ensuring that the $V$ curve in $\IV$ remains below the $V\rna$ curve. The latter constraint had no impact because the total and infectious viral titre remaining after the inoculum rinse at $t=0$ in the SC infection provides the most stringent constraint on $\PVest$ in the ED assay ($\le0.174$, see Methods). This constrains $\rho$ and $\gamma$ to relatively narrow 95\% CIs, despite their correlation (Figure \ref{fig:varyunits}H,I).

Wider credible intervals (95\%CI) for $\PVtoI$ and $\burst$ are obtained when $[V]=\IV$ compared to $[V]=\SIN$ (Figure \ref{fig:varyunits}K,L), yet this results in a somewhat better constrained $R_0$ (Figure \ref{fig:varyunits}M). A slightly smaller $R_0$ and $\tinf$ are estimated when $[V]=\IV$ rather than $\SIN$ (Table \ref{tab4pars},\ref{tab4pvals}), but these differences are unlikely to be biologically meaningful given typical inter-experimental variability \cite{paradis15}.

The establishment probability ($\PVest$), a biologically interesting quantity, can be computed from infection parameters when $[V]=\IV$ (it is identical to $\PVtoI$, Figure \ref{fig:varyunits}K) but not when $[V]=\SIN$. This is because it is impossible to know the relation between measured infections ($\SIN$) and infectious virions ($\IV$) from $\SIN$ alone, and consequently parameters in $\SIN$ cannot be converted to $\IV$. This constitutes yet another advantage of expressing MM variables and parameters using $[V]=\IV$.

\subsection{ODE versus stochastic models of infection}

The likelihood function $\LikeED$ is derived from the SM, and while it is deterministic, it accounts for the random nature of the infections taking place or not in the ED assay (shown in Figure \ref{inf-params-corrected}). Beyond this, results presented so far rely on the ODE to simulate the time course of the SC and MC infections. Now, the effect of using either the ODE or stochastic model (Eqn.\ \eqref{mfeq} in Methods) to simulate the SC and MC infections on the parameter estimation and the MM-predicted infection time course is explored, using the $\LikeED$ likelihood with $[V]=\IV$.

One challenge in estimating parameters using a SM is that any one parameter set randomly gives rise to different SM-predicted infection time courses, each with a different likelihood given the experimentally measured data. We circumvent this issue by adding the random number seed (\RNS) to the parameter set to be estimated by the SM compared to the ODE (${\pset}_\text{SM} = {\pset}_\text{ODE} \cup \{\RNS\}$). The \RNS is used to initialize the pseudorandom number generator at the start of the SM simulation such that any one SM parameter set (${\pset}_\text{SM}$) corresponds to a single, deterministic infection time course with its associated likelihood. This is not without cost: it adds an additional dimension to the parameter space the MCMC process must explore, one where the likelihood is jagged, i.e.\ small changes in the \RNS result in random, non-monotonic changes in $\LikeED$. In fact, small changes in the \RNS can result in large fluctuations in certain areas of the parameter space, e.g.\ $V(0)\sim\unit{1}{IV}$ or $R_0\lesssim1$. The cost of this additional parameter space dimension is in addition to the higher computational cost of numerically solving the SM over the ODE. Altogether, obtaining the final parameter set for the SM took $\sim15\times$ longer than for the ODE (27.5 days vs 1.79 days, see Methods). Nonetheless, adding the $\RNS$ to the SM parameters to be estimated offers a simple, practical solution to parameter estimation when using a SM.

\begin{figure}
\begin{center}
\includegraphics[width=1.0\linewidth]{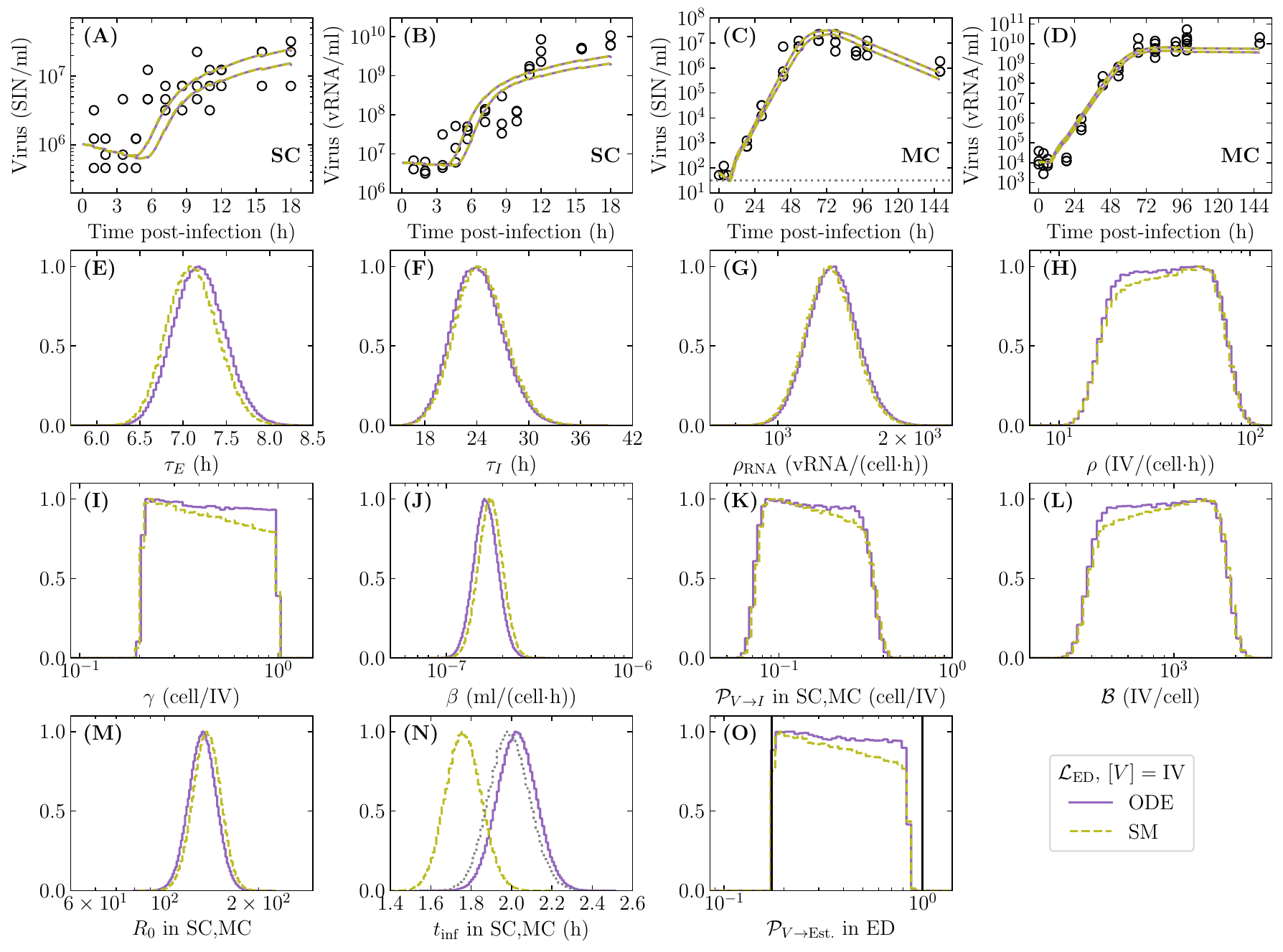}
\end{center}
\caption{%
\textbf{Impact of the random nature of infections on parameter estimation.} %
MM-predicted viral time courses and MPDs using the $\LikeED$ with $[V]=\IV$ for the ODE model (ODE, solid purple) or stochastic model (SM, dashed olive).
Panels (A--N) are as described in Figure \ref{fig:varylike}, except for (L) $\burst$, (M) $R_0$, and (N) $\tinf$ which, for the SM, correspond to the average of the distribution of values these quantities can take for a given $\pset$. The average of the $\burst$ and $R_0$ SM distributions matches the ODE model's expressions, but this is not the case for $\tinf$ (${\tinf}^\mathrm{SM} = \sqrt{\pi}/2 \cdot {\tinf}^\mathrm{ODE}$, see Methods). In (N) average $\tinf$ in the SM (${\tinf}^\mathrm{SM}$, dahsed olive) and the $\tinf$ obtained when SM parameters are used directly in the ODE model's expression (${\tinf}^\mathrm{ODE}$, dotted grey) are shown.
(O) $\PVest$ is the probability that an infection starting from \unit{1}{\IV} will become established, given the cell density in the ED assay ($N_\text{cells,ED}/\Vinoc$), used to convert $\IV$ to $\SIN$ as per Eqn.\ \eqref{conv}.
}
\label{fig:varySM}
\end{figure}

Figure \ref{fig:varySM} shows that the SM yields viral time courses nearly indistinguishable from those of the ODE, and only very slight (insignificant) changes in the estimated parameters' MPDs, except for $\tinf$, the average time it takes for a newly infectious cell to infect its first cell, when neglecting infectious virus lost to decay or cell entry. While $\tinf=\sqrt{2/(\rho \cdot \gamma \beta N_\text{cells}/\Vol)}$ in the ODE model for any one parameter set $\pset$, the same $\pset$ in the SM yields a distribution of $\tinf$ values whose average is $\sqrt{\pi}/2 \approx 89\%$ that of the ODE model's value (see Methods), i.e.\ for identical parameters the infecting time in the ODE model is longer than the SM's average infecting time. Importantly, with the exception of $\tinf$, differences in the parameter posteriors estimated by the ODE and SM are all statistically insignificant (Table \ref{tab4pvals}), are smaller than those seen for the different likelihood functions and the units used to express $V$, and are unlikely to be biologically meaningful. This suggests that random effects play a negligible role over the course of the experimental in vitro virus infections studied here. This is comforting in light of recent work showing that, at least in some parameter regimes, stochasticity can result in established infection time course curves that have the same shape but are markedly shifted in time relative to the ODE-predicted curves \cite{sazonov20,morris24} (see also Figure \ref{tshift}).

\begin{figure}
\begin{center}
\includegraphics[width=\linewidth]{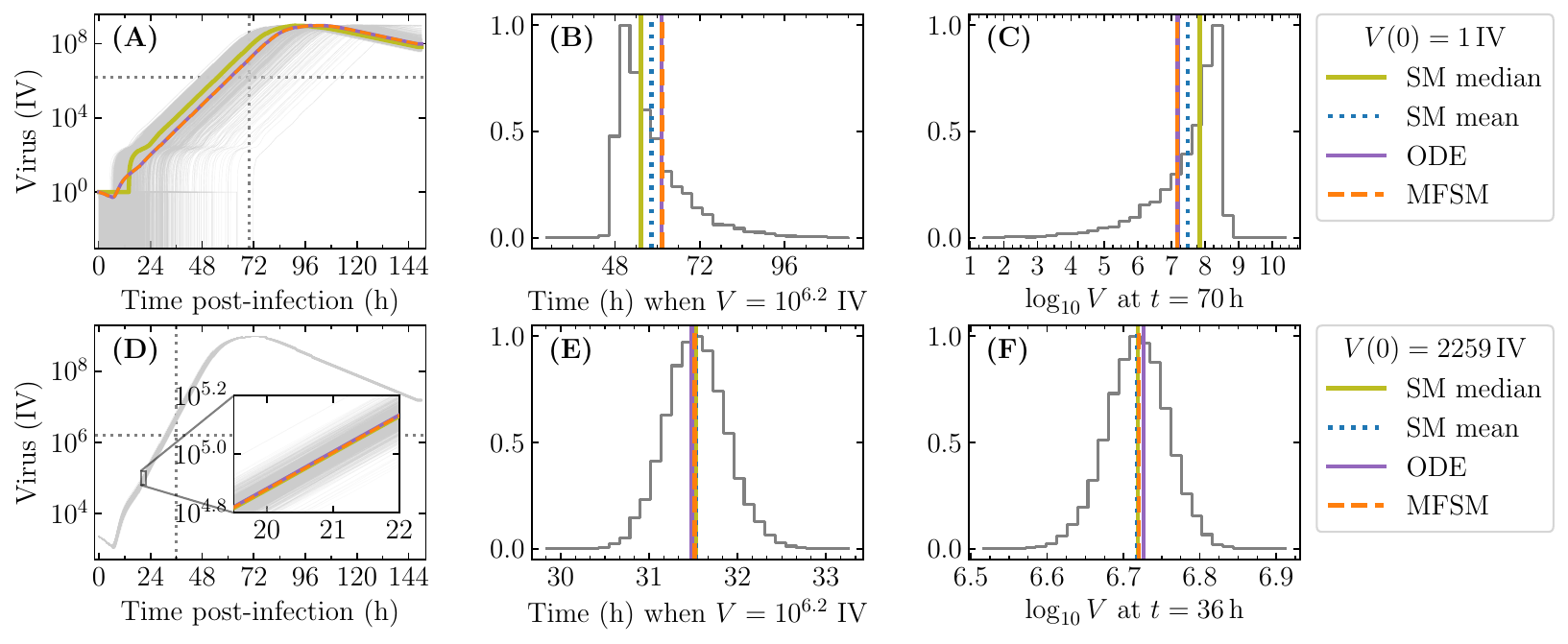}
\end{center}
\caption{%
\textbf{Effects of stochasticity on the distribution of infection time courses for different initial inocula.}
(A,D) Number of infectious virions (IV) as a function of time for an infection initiated with (A,B,C) \unit{1}{\IV} or (D,E,F) \unit{\mapVOmc}{\IV}, the MAP initial inoculum in the MC infection experiment. The individual trajectories of 10\,000 stochastic realizations that led to established infections (having discarded extinct infection), each initiated with a different random number seed, are shown (solid pale grey) along with their median (SM median) and mean (SM mean). The infection time course obtained with the ODE model and the mean-field version of the SM (MFSM), both using the SM's MAP parameter set (Table \ref{tab4pars}) are shown for comparison. 
(B,E) The distribution of times at which the virus load curves pass \unit{10^{6.2}}{\IV}, indicated as a horizontal dashed line in (A,D). 
(C,F) The distribution of infectious virions ($\log_{10} V$) at the specified time ($\unit{70}{\hour}$ or $\unit{36}{\hour}$) post-infection, indicated as a vertical dashed line in (A,D).
}
\label{stochMC}
\end{figure}

Figure \ref{stochMC} shows the infection time course predicted by the ODE model, the mean-field version of the SM (MFSM, see Methods), and 10,000 realizations of the SM, for established infections initiated with either a single infectious virion ($V(0) = \unit{1}{\IV}$, top row) or for the MAP initial inoculum in the low MOI, multi-cycle infection (\unit{\mapVOmc}{\IV}, bottom row), based on the experimental initial inoculum ($\unit{620}{\SIN}$, Table \ref{tab:fixdpars}) and the MAP $\PVest$ (Table \ref{tab4pars}). With a single infectious virion, the probability of infection extinction is high ($[1-\PVest]^{V_0=\unit{1}{\IV}} \sim 72\%$) and there is a noticeable time shift between the median and mean of the SM-predicted established infections and the ODE and MFSM solutions (Figure \ref{stochMC}A--C). With the estimated initial inoculum of \unit{\mapVOmc}{\IV} in the MC experiment, the probability of infection extinction is very low ($[1-\PVest]^{\unit{\mapVOmc}{\IV}}\approx 10^{-320}$), and the SM-predicted established infections are normally distributed (Figure \ref{stochMC}E,F) about their mean and median, and in agreement with ODE and MFSM solutions. Indeed, experimental infections, even those conducted at a low MOI, are designed to robustly result in established infections, i.e.\ the initial inoculum ($V(0)$) is chosen so that $[\PVext]^{V_0}\approx0$, to avoid wasteful, failed experimental infections. So while the parameters estimated herein can lead to infections in which significant random effects could be observed (Figure \ref{stochMC}A), by design such effects are not expected to play a meaningful role in experimental infections because of the chosen inoculum sizes.

\begin{figure}
\begin{center}
\includegraphics[width=1.0\linewidth]{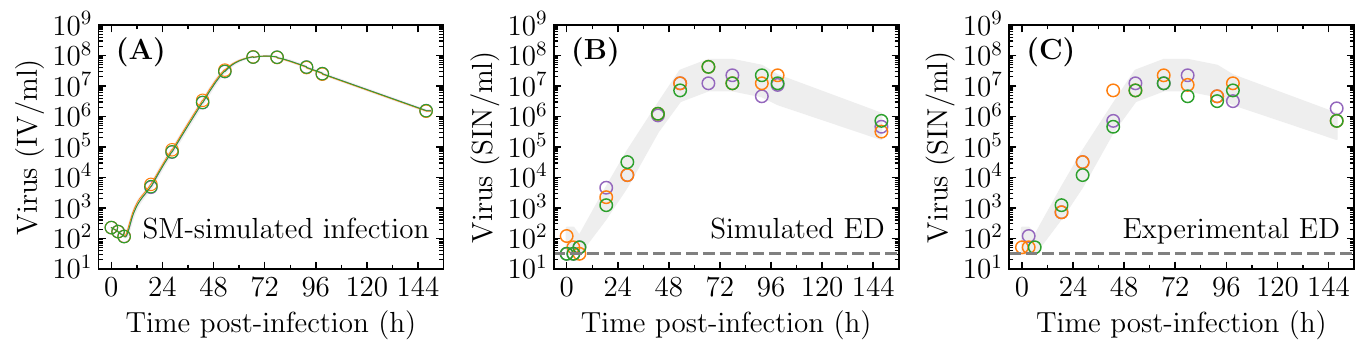}
\end{center}
\caption{%
\textbf{Sources of noise in experimental infections.} %
(A) SM-predicted $\IV$ concentration over time for the MC experiment, based on the MAP parameters, simulated in triplicate by using three different random number seeds ($\RNS$) and evaluated at each experimental measurement time (circles). The grey band, hardly visible, represents the 95\% CI of the $\IV$ concentration based on 10,000 SM-simulated infections (shown in Figure \ref{stochMC}D). %
(B) Simulated $\SIN$ concentration over time obtained by simulating an ED assay  for each $\IV$ concentration data point in (A). The grey band represents the 95\% CI of the $\SIN$ concentration at each time point, obtained by simulating 10,000 ED assay outcomes for the SM-predicted $\IV$ concentration at each experimental measurement time, based on the MAP parameters. %
(C) Experimentally measured $\SIN$ concentration over the course of the experimental MC infections performed in triplicate (circles) shown against the simulated ED 95\% CI from (B) for comparison. %
Infectious titers in $\SIN/\milli\litre$ below the lower limit of detection (dashed grey, Table \ref{tab:fixdpars}) are placed on the line. 
}
\label{noise}
\end{figure}

The relatively small variation amplitude among the 10,000 SM-predicted solutions, however, is at odds with the larger inter-replicate variability commonly observed in measured titres over the course of experimental infections (Figure \ref{fig:varySM}C vs Figure \ref{stochMC}D). Figure \ref{noise}A shows SM-simulated triplicate $\IV$ titre curves evaluated at each sampling time, using the MAP parameter set for the SM (Table \ref{tab4pars}) where the three replicates were generated by using three different random number seeds ($\RNS$). As in Figure \ref{stochMC}D, very little variation is observed in Figure \ref{noise}A between the SM-simulated triplicate $\IV$ concentrations at each sampling time, inconsistent with the experimentally observed variability (Figure \ref{fig:varySM}C). Next, each $\IV$ data point in Figure \ref{noise}A is used to simulate one realization of a stochastic ED assay, as described earlier (Figure \ref{algo}), to obtain a corresponding data point in $\SIN/\milli\litre$ (Figure \ref{noise}B). This process results in noticeable variability between the simulated $\SIN$ measurements at each sampling time. Indeed, the level of variability due to the ED assay is consistent with that observed experimentally (Figure \ref{noise}A vs.\ C). This suggests that the ED assay accuracy alone is sufficient to explain the variability observed in the experimentally measured titers, and that other factors that could affect inter-replicate variability, such as stochastic variability, are possibly minor and likely negligible compared to ED assay variability.

%%%%%%%%%%%%%%%%%%%%%%%%%%%%%%%%%%%%%%%%%%%%%%%%%%%%%%%%%%%%%%%%%%%%%%%%%%%%%%%%
%%%%%%%%%%%%%%%%%%%%%%%%%%%%%%%%%%%%%%%%%%%%%%%%%%%%%%%%%%%%%%%%%%%%%%%%%%%%%%%%
%%%%%%%%%%%%%%%%%%%%%%%%%%%%%%%%%%%%%%%%%%%%%%%%%%%%%%%%%%%%%%%%%%%%%%%%%%%%%%%%
\section{Discussion}

The course and outcome of experimental virus infections are tracked via repeated sampling over time, and subsequent measurement of the virus concentration in the samples. In particular, a sample's virus infectivity is measured using an infectivity assay, such as a plaque or focus forming assay or a \tcid assay, which counts the number of infections the virus-containing sample can cause per unit volume. But some or even all of the infectious virions contained in the inoculant of an infectivity assay's wells can randomly fail to cause or to establish an infection, go uncounted, and hence affect the resulting measured infection concentration. The stochastic nature of infections in either the infection experiments itself or in the measurement of its samples via an infectivity assay, could have an important effect on our interpretation of the infection time course and on the mathematical estimation of its infection parameters.

By simulating a \tcid endpoint dilution (ED) assay, we demonstrated that the mathematical parameters of the infection determine the probability that an infectious virion ($\IV$) contained within a sample will result in an established infection, and thus be counted as an infection-causing dose (or specific infection, $\SIN$) by the assay. Specifically, we found that this probability ($\PVest$) decreases when decreasing the probability of a successful cell infection post viral entry ($\gamma$) or, to a lesser degree, the ratio between loss of virus due to cell entry and that due to loss of infectivity ($(\beta N_\text{cells}/\Vol)/c$). In contrast, the average total number of infectious virions released by an infected cell over its infectious lifespan (mean burst size, $\mathcal{B} = \rho \cdot \tau_I$) and the shape parameter of the Erlang-distributed time spent by infected cells in the infectious phase ($n_I$), quantities that can have an effect on the extinction probability in principle, had a negligible effect. This is mostly because, in contrast to $\gamma$ and $(\beta N_\text{cells}/\Vol)/c$, $\mathcal{B}$ and $n_I$ affect the likelihood of infection extinction only after the successful infection of a cell, whose infectious virus progeny is typically sufficient to ensure infection establishment. We demonstrated that the experimentally measured infection causing dose concentration in a sample (e.g., $\SIN/\milli\litre$) corresponds exactly to the product of the actual infectious virion concentration ($\IV/\milli\litre$) and the establishment probability ($\PVest$ or $1-\PVext$) of an infection initiated with one such infectious virion. By affecting the establishment probability, the infection parameters also affect the observed time course of experimentally measured $\SIN/\milli\litre$ over the course of the infection, the very data that is used to estimate their value.

Thus, we next considered whether explicitly and correctly accounting for the parameters' effect on experimental titer measurements, on which parameters depend for their estimation, would hinder or even preclude parameter estimation altogether. This was achieved in three stages, and parameter estimation was performed at each stage using experimental data from influenza A virus infection experiments previously analyzed \cite{simon16}. Firstly, we considered the effect of the likelihood function, i.e.\ the metric by which we measure agreement between the mathematical model's prediction and experimental measurements. We compared the typical likelihood function which assumes the experimental concentration of $\log_{10}$ $\SIN$ in a sample follows a normal distribution ($\LikeNR$) versus a novel expression that correctly computes the likelihood of having experimentally observed the number of wells infected at each dilution of the ED assay given the ODE-predicted number of $\SIN$ in that sample ($\LikeED$). Secondly, we compared the effect of expressing the variable representing infectious virus in the ODE model directly using the experimental measurement units ($[V]=\SIN$) versus using the more physically meaningful units of infectious virions ($[V]=\IV$), and then using the model's parameters, and their corresponding $\PVest$, to convert the ODE-predicted $\IV$ into $\SIN$ for use in the new likelihood function ($\LikeED$). Thirdly, we compared the effect of the typical ODE model versus using a stochastic model (SM) to represent the course of the in vitro infections.

Changing the likelihood function ($\LikeNR$ vs $\LikeED$) had the largest impact on the estimated parameters, shifting the average value of all estimated parameters. However, none of the changes were statistically significant (Table \ref{tab4pvals}). An important advantage of the new $\LikeED$ is that it correctly and trivially handles measurements below the ED assay's limit of detection, i.e., when, at all dilutions of the performed ED assay, either no well is infected or all are.

Changing the units by which infectious virus is expressed in the ODE model ($[V] = \SIN$ vs $\IV$) had no effect on half of the parameters --- the duration of the eclipse ($\tau_E$) and infectious ($\tau_I$) phases of an infected cell's lifespan, and the total virus production rate per cell ($\rho\rna$) --- but drastically changed the posterior distribution for the estimated rate of infectious virus production ($\rho$), the rate of virion loss due to cell entry ($\beta$), and the rate of cell infection per virus entry into cell ($\gamma$). This is because the physical meaning of these three parameters ($\rho$, $\beta$, $\gamma$) changes when the units used to represent virus in the ODE model change. For example, $\rho$ is the number of $\SIN$ produced per cell per hour when $[V]=\SIN$ but it is the number of $\IV$ produced per cell per hour when $[V]=\IV$. Importantly, when $[V]=\SIN$, experimental measurements (total and infectious viral titer) constrain the value of $\rho$ and of the product ($\gamma\beta$), and impose a lower bound on $\gamma$ (upper bound on $\beta$). The measurements do not, however, provide an upper bound on $\gamma$ (lower bound on $\beta$), compromising the estimation of these two parameters. This is because one $\SIN$ (one observed infection) can correspond to multiple infectious virions, allowing more than one cell to become infected per $\SIN$ cell entry ($\gamma>\unit{1}{cell/\SIN}$). In contrast, when $[V]=\IV$, experimental measurements constrain the value of $\beta$ and of the product ($\gamma\rho$), and measurement of total virus (viral RNA) imposes a lower bound on $\gamma$ (upper bound on $\rho$) since there cannot be more infectious virions ($\IV$) than there are virions counted in total (viral RNA). This highlights the critical importance of tracking both infectious (SIN) and total (vRNA) virus to properly constrain these parameters. A physical upper bound can be imposed on $\gamma$ (lower bound on $\rho$) since entry of one infectious virion into a cell can at most result in the infection of one cell ($\gamma\le\unit{1}{cell/\IV}$), and allowing for the individual estimates of $\gamma$ and $\rho$ to be well constrained. This is an important advantage of expressing $[V]$ in infectious virions.

Another important gain in expressing virus in units of infectious virions is that it yields physically meaningful quantities. Experimental measurements of a virus sample's infectivity in units such as \tcid, $\SIN$, PFU, or FFU, represent a count of infections caused, not a count of the number of virions physically capable of causing infection, whether they go on to do so or not. Indeed, while the former is a shared, combined property of the virion, the cells, and experimental conditions such as temperature, medium used, inoculum incubation time, etc., i.e., it is specific to the infection conditions, the latter is an absolute property of the virion alone, independent of all other factors. The ODE and stochastic models consistently estimated that \unit{1}{\SIN} corresponds to 1.1 to 5.8 infectious virions (95\% CI of $1/\PVest$ in ED), under the conditions of the experiments analyzed herein. Over the course of its infectious lifespan, we estimate that one A549 cell produced on average a total of 400 to 2\,000 infectious virions (95\% CI of $\burst$). This quantity, i.e.\ the actual number of infectious virions rather than the conditions-dependent infectious dose, has been computed previously, but typically has not included the probability of failure of infection post virus entry into a cell ($\gamma$), and has relied on a guess as to the infectious virion production rate ($\rho$) and the rate of cell entry by infections virions ($\beta$), e.g.\ \cite{pearson11,conway13,czuppon21}. For example, in Pearson et al.\ \cite{pearson11}, the average burst size ($\rho \cdot \tau_I$), hence $\rho$, is guessed to be $\unit{10}{IV/cell}$ and $\beta$ is chosen in order to have a ``sensible'' value for the basic reproductive number ($R_0$), which is computed using the guessed burst size value. We found that it takes on average 1 to 5 infectious virion cell entry for one successful cell infection (95\% CI of $1/\gamma$). To our knowledge, this is the first time that this quantity has been estimated. It is consistent with an earlier report that approximately 50\% of influenza A virus cell entry fail to result in successful fusion of the virion with its endosome \cite{stegmann93,heldt12}. This would correspond to 2 infectious virion cell entry per successful cell infection, but it could be less if some non-infectious virions are capable of cell entry, and it could be more if other steps than fusion contribute to infection failure of otherwise fully infectious virions, which are both likely.

Changing from an ODE to a stochastic model to simulate the course of the infection had the smallest impact out of the three stages, despite its important increase (15-fold) in the computational cost required to estimate the parameters. The amount of variability observed with the stochastic model is largest, and its solutions deviate most from those of the ODE model, when the infection's establishment probability is low. Since experimental virus infections in vitro, even those inoculated with a relatively low MOI, are designed to ensure infection will take hold, it is not surprising that our estimated establishment probability was essentially 100\%, resulting in insignificant stochastic fluctuations per the stochastic model's predictions. Yet the low level of stochasticity predicted by the SM's solutions was inconsistent with the much greater amount of variability observed between infection replicates. When the SM-predicted number of infectious virions over the course of the infection was used in simulated stochastic ED assays, the resulting estimated SIN concentration at each of the simulated sampling time points was consistent with the experimental inter-replicate variability. These findings suggest that inter-replicate noise in infection experiments is introduced primarily by the infectivity measurement assays, such as the ED assay. This means inter-replicate variations could be reduced significantly by improving ED assay accuracy, e.g.\ by decrease the dilution factor or increasing the number of replicates per dilution of the ED assay \cite{cresta21}.

There are a number of limitations to the analyses and results reported herein. Firstly, we assumed that one viral RNA, as measured via qRT-PCR, equals one virion. This assumption, which is highly dependent on having a well-validated standard curve, affects the lower bound constraint on $\gamma$ when $[V]=\IV$ because we require that there be no more infectious virions than there are vRNAs. Herein, this constraint was imposed by a single pair of points: the total (vRNA) and infectious ($\SIN$) viral titers measured after the inoculum rinse at $t=0$ in the single-cycle infection experiment (see Methods). Secondly, we assumed that the state of the virions when they are sampled from the infection supernatant is the same as that when they were inoculated into the ED assay wells, i.e.\ the number of $\IV/\milli\litre$ is unchanged. Most likely, a fraction of the virions sampled from the infection experiments lost their infectivity --- either prior to freezing, as a result of being frozen, or post-thaw --- by the time they were inoculated onto the ED assay wells to be measured. It has been reported that there can be a 10-fold reduction in the number of infections a sample will cause, measured by plaque assay, after each freeze/thaw cycle \cite{cbeau08}. This means that there were more IV in the infection experiment than were counted in the ED assay. This could easily be remedied by performing two ED assays for a few samples: one immediately after they were harvested from the infection experiment and another after the sample was frozen and thawed, as per normal protocols. By comparing these two $\SIN$ measures, one could correctly account for the proportion of virion infectivity loss associated with the more convenient and common protocol. This assessment would have to be repeated whenever experimental conditions change, e.g.\ for different cell types or virus strains or if different buffers are added prior to freezing samples.

Thirdly, and perhaps most critically, we had to assume that the ED assay and infection experiments were conducted under the same conditions, and as such shared all infection parameters ($\beta$, $\gamma$, $\rho$, $\tau_E$, etc.). This was not the case: infections were conducted in A549 cells while ED assays were performed in MDCK cells. This implies that, for example, one SIN in A549 cells could correspond to \unit{10}{IV}, whereas one $\SIN$ in MDCK cells could represent \unit{2}{\IV} if the establishment probability of one IV in MDCK cells was 5 times greater than in A549 cells. This assumption was necessary because the ED assay alone is insufficient to estimate the infection parameters in MDCK cells, required to estimate the infection establishment probability in the ED assay, which would differ from that in A549 cells.

But performing the infection experiments and the ED assay in different cell types is not only a concern for the methodology introduced herein, but more generally will result in misleading interpretations of virological data. Consider two viruses: virus A is more infectious in MDCK cells (more SIN/IV in the ED assay) than in A549 (fewer infections/IV in the infection experiment), but the opposite is true for virus B. If virus A and B are inoculated at equal MOI based on their ED assay-assessed infectivity, an equal number of SIN of virus A and B based on the ED assay will result in an actual infection MOI that is smaller for virus A than for virus B, i.e.\ virus A will cause the infection of possibly far fewer cells (at supposedly equal MOI). This will make virus A appear to take more time to reach peak titer (because it caused fewer initial infections), but will make it appear as though it reaches a higher peak titer (because titer is measured in MDCK cells). For these reasons, any future work should ensure that the ED assay and infection experiments are conducted in the same cells, under the same conditions (temperature, medium content, etc.), except for the number of cells and total inoculum volume which can be accounted for mathematically, as was done herein.

In conclusion, based on the analysis herein, we would recommend that future work: (1) measure both total and infectious virus titer over the course of experimental infections; (2) make use of the \tcid ED assay to quantify virus infectivity; (3) that the ED assay and infections be performed in the same cells and under similar experimental conditions; (4) that parameter estimation be performed using the physically accurate likelihood function introduced herein ($\LikeED$); and (5) that units of infectious virions rather than infectious dose (e.g.\ \tcid, $\SIN$, PFU, FFU) be used to express virus in the mathematical model. We do not, however, recommend the use of a stochastic model to simulate the course of experimental infections in vitro because its higher computational cost did not translate to any meaningful differences in the estimated parameters.

%%%%%%%%%%%%%%%%%%%%%%%%%%%%%%%%%%%%%%%%%%%%%%%%%%%%%%%%%%%%%%%%%%%%%%%%%%%%%%%%
%%%%%%%%%%%%%%%%%%%%%%%%%%%%%%%%%%%%%%%%%%%%%%%%%%%%%%%%%%%%%%%%%%%%%%%%%%%%%%%%
%%%%%%%%%%%%%%%%%%%%%%%%%%%%%%%%%%%%%%%%%%%%%%%%%%%%%%%%%%%%%%%%%%%%%%%%%%%%%%%%
\section{Methods}

\subsection{Mathematical models}
\label{mathmod}

The ordinary differential equation (ODE) model and its stochastic model (SM) counterpart used herein are given by
\begin{align}
\dift{T} &= -\gamma \beta_\text{} TV_\text{}/\Vol &T^{t+1} &= T^t-N^{\text{inf}} \nonumber \\
\dift{E_1} &= \gamma \beta_\text{} TV_\text{}/\Vol - \frac{n_E}{\tau_E} E_1 &E_1^{t+1} &= E_1^t + N^{\text{inf}} - E_1^{\text{out}} \nonumber \\
\dift{E_i} &= \frac{n_E}{\tau_E}E_{i-1} - \frac{n_E}{\tau_E}E_i &E_i^{t+1} &= E_i^t + E_{i-1}^{\text{out}} - E_i^{\text{out}} \qquad i=2,3,...,n_E \nonumber \\
\dift{I_1} &= \frac{n_E}{\tau_E}E_{n_E}-\frac{n_I}{\tau_I}I_1 &I_1^{t+1} &= I_1^t + E_{n_E}^{\text{out}} - I_1^{\text{out}} \label{mfeq} \\
\dift{I_j} &= \frac{n_I}{\tau_I}I_{j-1}-\frac{n_I}{\tau_I}I_j &I_j^{t+1} &= I_j^t + I_{j-1}^{\text{out}} - I_j^{\text{out}} \qquad j=2,3,...,n_I \nonumber \\
\dift{V} &= \rho_\text{}\sum_{j=1}^{n_I}I_j-cV_\text{}-\beta_{} TV_\text{}/\Vol &V^{t+1} &= V^t + V^\text{prod} - V^\text{decay} - V^\text{enter} \nonumber \\
\dift{V_\text{RNA}} &= \rho_\text{RNA}\sum_{j=1}^{n_I}I_j-c_\text{RNA}V_\text{RNA}-\beta_{} TV_\text{}/\Vol &V_\text{RNA}^{t+1} &= V_\text{RNA}^t + V_\text{RNA}^\text{prod} - V_\text{RNA}^\text{decay} - V^\text{enter} \nonumber
\end{align}
and are largely identical to those introduced and used in \cite{quirouette23}. The SM is similar to the ODE model, where SM variables denoted with a superscript $t$ are whole numbers, e.g.\ $T^t$ represents the discrete number of target cells at time $t$. The remaining terms in the SM, corresponding to changes in the SM variables, are random whole numbers, generated at each time step as described in Table \ref{summary-rvar}. $V^\text{prod}$ is drawn from $\text{Binomial}(n=V_\text{RNA}^\text{prod},\,p_\pi=\rho/\rho_\text{RNA})$ rather than $\text{Poisson}(\lambda=\Delta t\cdot \rho\sum_{j=1}^{n_I} I_j^t)$ as in \cite{quirouette23} so that, at each time step, there is never more infectious than total virions produced. The total number of virions, $V\rna$, is expressed in units of vRNA and variable $V$ corresponds either to the number of infection-causing doses (SIN) or infectious virions (IV). We further consider the mean-field solution to the SM (MFSM) wherein the distributions for the SM's random variables in Table \ref{summary-rvar} are replaced by their average. In the ODE, MFSM, and SM, cells and viruses are counts (e.g., $T = 1.3$ cells in ODE and MFSM, and either 1 or 2 in the SM) rather than concentrations.

\begin{table*}
\caption{Random variables of the SM.}
\label{summary-rvar}
\begin{center}
\begin{tabular}{ll}
Random variable & Random number generator \\
\hline
$E_i^\text{out}$ & $\text{Binomial}(n=E_i^t,\, p_E=\Delta t\cdot n_E/\tau_E)$ where $i=1,2,...,n_E$\\
$I_j^\text{out}$ & $\text{Binomial}(n=I_j^t,\, p_I=\Delta t\cdot n_I/\tau_I)$ where $j=1,2,...,n_I$\\
$V^\text{prod}$ & $\text{Binomial}(n=V_\text{RNA}^\text{prod},\,p_\pi=\rho/\rho_\text{RNA})$ \\
$V_\text{RNA}^\text{prod}$ & $\text{Poisson}(\lambda=\Delta t\cdot \rho_\text{RNA} \sum_{j=1}^{n_I} I_j^t)$ \\
$V^\text{decay},V^\text{enter},V^\text{remain}$ & $\text{Trinomial}(n=V^t,\, p_1=\Delta t\cdot c,\, p_2=\Delta t\cdot\beta T^t/\Vol,\, p_3=1-p_1-p_2)$ \\
$V^\text{decay}_\text{RNA}$ & $\text{Binomial}(n=V_\text{RNA}^t,\, p_d=\Delta t\cdot c_\text{RNA})$ \\
$N^\text{inf}$ & $|\{x_i | x_i \in \mathcal{U}\{a=1,\, b=T^t\}, 1 \leq i \leq V^\text{suc}\}|^\dagger$ \\
& \hspace{2em} where $V^\text{suc} = \text{Binomial}(n=V^{\text{enter}},\,p_V=\gamma)$ \\
\hline
\end{tabular}
\begin{minipage}{0.80\linewidth}
$^\dagger$ $N^\text{inf}$ is equal to the cardinality, i.e.\ the number of unique elements, of the set of $V^\text{suc}$ random numbers drawn from the discrete uniform distribution over the interval $[0,T^t]$.
\end{minipage}
\end{center}
\end{table*}

In numerically solving the SM, the duration of the discrete time steps, $\Delta t$, are computed at each iteration step $t$ as
\begin{align}
\Delta t = \frac{p_\text{events}}{
\max \bigg\{ \frac{\beta T^t}{\Vol}, c, c_\text{RNA}, \frac{n_E}{\tau_E}, \frac{n_I}{\tau_I} \bigg\} }
\end{align}
where $p_\text{events}=0.05$ (or 5\%) is the probability of occurrence of the most likely event (see Table \ref{summary-rvar}), which was shown in \quirouette to correspond to sufficiently small steps so as to provide an accurate solution.

An important feature of the SM is that an infection can become extinct rather than established, i.e.\ it can randomly fail to take hold or spread significantly. The extinction probability of an infection given that there is initially one infectious virion, $\PVext$, was derived in \cite{quirouette23} for this SM and is given in Eqn.\ \eqref{PVext}. The establishment probability of an infection given that there is initially one infectious virion, $\PVest$, is simply given by $1-\PVext$.

\subsection{Base infection parameters}
\label{mod}

The base infection parameters, used in Figures \ref{tshift}, \ref{inf-params} and \ref{inf-params-corrected}, were adapted from those reported in \simon for the seasonal influenza A virus strain A/Canada/RV733/2003 (a A/New Caledonia/20/1999-like clinical isolate), i.e.\ $\tau_E = \unit{7}{\hour}$, $n_E = 60$, $\tau_I = \unit{41}{\hour}$, $n_I = 60$, $c = \unit{0.0573}{\hour^{-1}}$. The ODE model in \simon does not have separate terms to represent rate of virions loss to cell entry ($\beta T V/\Vol$) and rate of cell loss to infection ($\gamma\beta T V/\Vol$) and did not estimate $\gamma$. Unless otherwise stated, base values of $\gamma = \unit{1}{cell/\IV}$ and $\Vol = \unit{1}{\milli\litre}$ were used. Additionally, infection parameters that include units of tissue culture 50\% infectious dose (\tcid) were modified to have units of infectious virions (\IV) by assuming that $\unit{1}{\tcid} \approx \unit{0.56}{SIN}$ \cite{wulff12} and $\unit{1}{SIN} \approx \unit{1}{\IV}$ for simplicity. Parameter $\beta_{\tcid}$ in \simon corresponds roughly to $\gamma\beta/\Vol$ in our SM and thus was set to $\beta_{\tcid} \cdot \Vol \cdot (\unit{1}{\tcid}/\unit{0.56}{\SIN}) \cdot (\unit{1}{\SIN}/\unit{1}{\IV}) = \unit{10^{-5.0}}{\milli\litre/(cell \cdot \hour)}$. The infectious virus production rate per volume in \simon, $\rho_{\tcid}$, corresponds to $\rho\cdot\Vol$ in our SM and thus was set to $\rho_{\tcid}/\Vol \cdot (\unit{0.56}{\SIN}/\unit{1}{\tcid}) \cdot (\unit{1}{\IV}/\unit{1}{\SIN}) = \unit{10^{0.86}}{\IV/(cell \cdot \hour)}$.

\subsection{Simulations of endpoint dilution (ED) assay experiments}
\label{sim-ED-assay}

The process of simulating an ED assay outcome, illustrated in Figure \ref{algo}, was used to produce Figures \ref{inf-params}, \ref{inf-params-corrected}, and \ref{noise}B. A simulated ED assay experiment for a virus sample with an actual infectious virion concentration $C_\text{actual}$ begins by determining the random number of $\IV$ deposited in each well of replicate row $i$ in dilution column $j$, $V_0^{i,j}$. These random numbers are drawn from $\text{Binomial}(n_j\,=\Vinoc \mathcal{D}_j/V_\text{vir},\,p\,=\,C_\text{actual}V_\text{vir})$ where $\Vinoc$ is the total volume of inoculum placed in each well, $\mathcal{D}_j\in(0,1]$ is the dilution factor for column $j$, and $V_\text{vir} = \unit{5.236 \times 10^{-16}}{\milli\litre}$ \cite{noda11} is the volume of a single influenza A virion (see \cresta for details). To determine whether any one well becomes infected, a uniform random number $r\in[0,1)$ is drawn, and the well is uninfected if $r<{(\PVext)}^{V_0^{i,j}}$, and infected otherwise. As a final step, the most likely $\log_{10}$ infectious dose concentration ($\log_{10}(\SIN/\milli\litre)$) is determined by providing the number of infected wells in each dilution column to the midSIN calculator \cite{cresta21}.

\subsection{Simulating the single- and multiple-cycle infections}

\begin{table}
\caption{Fixed initial conditions$^a$ and parameters of the MM and likelihood function}
\label{tab:fixdpars}
\begin{center}
\begin{tabular}{lll}
\hline
Parameter & Symbol & Value \\
\hline
Infectious virus clearance rate & $c$ & \unit{0.0573}{\perh} \\
Total virus clearance rate & $c\rna$ & \unit{0.001}{\perh} \\
\# of eclipse compartments & $n_E$ & 60 \\
\# of infectious compartments & $n_I$ & 60 \\
\# of cells in SC/MC infections & $\Ncells$ & \unit{1.9\times10^6}{cells} \\
Total volume in SC/MC infections & $\Vol$ & \unit{10}{\milli\litre} \\
Dilution due to sampling in SC/MC infections & $f_\text{sampling}$ & 0.95 \\
\hline
SC initial infectious titre pre-rinse & $V(-\unit{1}{\hour})_\text{SC}$ & based on MOI=3 \\
SC initial infectious titre post-rinse & $V(0)_\text{SC}$ & \unit{10^{7}}{\SIN}$^b$ \\
SC initial total titre pre-rinse & $V\rna(-\unit{1}{\hour})_\text{SC}$ & \unit{10^{10.0}}{vRNA} \\
SC initial total titre post-rinse & $V\rna(0)_\text{SC}$ & \unit{10^{7.76}}{vRNA} \\
\hline
MC initial infectious titre & $V(0)_\text{MC}$ & \unit{10^{2.79}}{\SIN}$^b$ \\
MC initial total titre & $V\rna(0)_\text{MC}$ & \unit{10^{5.02}}{vRNA} \\
\hline
$\Like\rna$'s SC std.\ dev.\ of $\log_{10}$ total titre & $\sigma_{V_\text{RNA,SC}}$ & 0.237 \\
$\Like\rna$'s MC std.\ dev.\ of $\log_{10}$ total titre & $\sigma_{V_\text{RNA,MC}}$ & 0.258 \\
$\LikeNR$'s SC std.\ dev.\ of $\log_{10}$ infectious titre & $\sigma_{V_\text{SC}}$ & 0.247 \\
$\LikeNR$'s MC std.\ dev.\ of $\log_{10}$ infectious titre & $\sigma_{V_\text{MC}}$ & 0.240 \\
$\LikeNR$'s lower limit of detection & $C_\text{LLoD}$ & \unit{10^{1.494}}{\SIN^b/\milli\litre} \\
ED assay total well volume & $\Vinoc$ & \unit{0.05}{\milli\litre} \\
ED assay \# of cells/well & $N_\text{cells,ED}$ & \unit{10^5}{cells} \\
ED assay \# of dilution columns & $\text{col}$ & 8 \\
ED assay \# of replicate per dilution & $n_r^\text{col}$ & 4 \\
ED assay dilution factor & $\Dil$ & 0.1 (10-fold) \\
\hline
\end{tabular}
\end{center}
$^a$ In the SM, initial conditions for cells and virus are rounded to the nearest integer. \\
$^b$ When $[V]=\IV$, Eqn.\ \eqref{conv} is used to convert $\SIN$ to $\IV$ based on the MM parameters.
\end{table}

The ODE or stochastic models in Eqn \eqref{mfeq} was used to simulate the experimental single-cycle (SC) and multiple-cycle (MC) infections described and analyzed in \simon, following a similar procedure. In the SC experiment, the virus was inoculated onto the cells at $t = \unit{-1}{h}$ to achieve a MOI of 3 within one hour, and the cell culture was rinsed at time $t = \unit{0}{h}$ to remove any virus still present from the high inoculum. The simulated SC infection was initiated at $t = \unit{-1}{h}$, with all cells considered uninfected and susceptible ($T(-\unit{1}{h})=\Ncells=\unit{1.9\times 10^6}{cells}$, and $E_i=I_j=0$). The initial number of vRNA, $V\rna(\unit{-1}{h})_\text{SC}$, was set to the single qRT-PCR total viral concentration measurement taken at that time (vRNA/\milli\litre) multiplied by the inoculum volume $\Vol=\unit{10}{\milli\litre}$. The initial number of $\SIN$ or $\IV$, $V(\unit{-1}{h})_\text{SC}$, was set so as to achieve the experimental multiplicity of infection ($\text{MOI} = 3$) within the \unit{1}{\hour} incubation time given the parameters ($\pset$), as follows. The expected fraction of uninfected cells for a given multiplicity of infection (MOI) is $\approx \me^{-\text{MOI}}$ \cite{liao16}. During the incubation period, which is shorter than the average length of the eclipse phase, one can assume that virus production is negligible. Therefore, the kinetics of target cells ($T$) and infectious titre ($V$) over this period of time can be represented by
\begin{align}
\dift{T} &= -\gamma \beta TV/\Vol \nonumber \\
\dift{V} &= - cV -\beta TV/\Vol \label{MOIode}
\end{align}
The initial amount of infectious virus $V(-\unit{1}{\hour})_\text{SC}$ to achieve a given MOI at $t=0$ can be determined by finding the root of the function that takes $V(-\unit{1}{\hour})_\text{SC}$ as an argument and returns the difference $\left[T(0)/\Ncells-\me^{-\text{MOI}}\right]$, where $T(0)$ is determined by numerically solving Eqn.\ \eqref{MOIode} with initial conditions $V(-\unit{1}{\hour})_\text{SC}$ which is to be determined and $T(-\unit{1}{\hour})_\text{SC} = N_\text{cells}$. The number of SIN and vRNA post-rinse at time $t = \unit{0}{\hour}$, $V(0)_\text{SC}$ and $V_\text{RNA}(0)_\text{SC}$, was set to the geometric mean of all experimental measurements from $\unit{0}{hpi}$ up to $\unit{3.5}{hpi}$, in $\SIN/\milli\litre$ or vRNA/\milli\litre, multiplied by $\Vol$. If the MM expresses $V(0)_\text{SC}$ in $\IV$, the value in $\SIN$ is converted to $\IV$ as per Eqn.\ \eqref{conv}, where $\PVest=(1-\PVext)$ is determined based on the MM parameters (Eqn.\ \eqref{PVext}).

The MC infection is initiated at time $t = \unit{0}{h}$ with no rinse thereafter, since the inoculum is small in order to have multiple rounds of cell infections, i.e.\ multiple cycles. At $t=0$, all cells are considered uninfected and susceptible ($T(\unit{0}{h})_\text{MC}=\Ncells=\unit{1.9\times 10^6}{cells}$, and $E_i=I_j=0$). The initial number of $\SIN$ and vRNA, $V(0)_\text{MC}$ and $V_\text{RNA}(0)_\text{MC}$, was set to the geometric mean of all experimental measurements up to $\unit{7}{hpi}$ in $\SIN/\milli\litre$ or vRNA/\milli\litre, multiplied by the volume of supernatant $\Vol = \unit{10}{\milli\litre}$. When $V(0)_\text{MC}$ is expressed in $\IV$, \SIN\ converted to $\IV$ using Eqn.\ \eqref{conv}, as in the SC simulation.

Over the course of both the SC and MC experimental infections, at each sampling time, $\unit{0.5}{\milli\litre}$ of the $\unit{10}{\milli\litre}$ supernatant is removed for virus quantification, and replaced with $\unit{0.5}{\milli\litre}$ of fresh media. In the ODE model, this dilution of the supernatant at each sampling time is implemented by multiplying $V$ and $V\rna$ by factor $f_\text{sampling} = 1-(\unit{0.5}{\milli\litre})/(\unit{10}{\milli\litre}) = 0.95$. In the SM, a binomial random variable with a probability of success of $p=f_\text{sampling}=0.95$ is used to determine how many $V$ and $V\rna$ particles will remain after the sampling.

As in \simon, the number of compartments for the eclipse and infectious phases were fixed so that time spent in the eclipse or the infectious phase follows a normal-like distribution \cite{holder11}, i.e.\ $n_E = n_I = 60$. The values of the rate of infectious and total viral decay were fixed to the values determined in \simon, $c = \unit{0.0573}{h^{-1}}$ and $c_\text{RNA} = \unit{0.001}{h^{-1}}$, based on a viral decay (mock-yield) assays performed therein. Table \ref{tab:fixdpars} summarizes all the fixed parameters and initial conditions used in simulating these infections.

\subsection{Measurements' likelihood and parameter estimation}

The posterior probability of a given parameter set, $\pset$, given the set of all total and infectious viral titre measurements over the course of the SC and MC infections, $\{\text{data}\} = \{\Cdata_\text{SC},\Cdata_{RNA,SC},\Cdata_\text{MC},\Cdata_\text{RNA,MC}\}$, is given by
\begin{align}
\mathcal{P}_\text{post}(\pset|\sdata,\text{MM})
	\propto
\Like_{V\rna}(\Cdata\rna|\pset,\MM)
\cdot \Like_V(...|...,\pset,\MM)
\cdot \prior(\pset)
\label{eqn:likepost}
\end{align}
where $\prior(\pset)$ is the prior probability of $\pset$ which includes any known physical constraints or any prior knowledge, and $\Like_{V\rna}$ and $\Like_V$ correspond to the likelihood of the observed experimental measurements of the total and infectious viral titre concentrations in the supernatant samples collected in triplicate at each time point over the course of the in vitro infections.

The likelihood of the total viral titre measured via qRT-PCR in units of vRNA/$\milli\litre$ in both the SC and MC infections is given by
\begin{multline}
\Like_{V\rna}(\Cdata\rna|\pset,\MM) =
\prod_{r=1}^3 \prod_{t=t_\text{SC}}
\exp\left\{
- \frac{\left[\log_{10} \Cdata_{\text{RNA,SC},r}(t) -\log_{10} C^\text{model}(t|\pset) \right]^2}{2\sigma^2_\text{RNA,SC}}
\right\} \\
\cdot \prod_{r=1}^3 \prod_{t=t_\text{MC}}
\exp\left\{
- \frac{\left[\log_{10} \Cdata_{\text{RNA,MC},r}(t) -\log_{10} C^\text{model}(t|\pset) \right]^2}{2\sigma^2_\text{RNA,MC}}
\right\}
\end{multline}
where $\Cdata_{\text{RNA,SC},r}(t)$ is the experimentally measured vRNA/\milli\litre, $t$ are the sampling times which are distinct in the SC and MC infections, $r=1,2,3$ is the infection replicate since both the SC and MC infections were performed in triplicate, and $C^\text{model}(t|\pset)=V\rna(t)/\Vol$ is the MM-predicted vRNA for the simulated SC or MC infection divided by $\Vol=\unit{10}{\milli\litre}$, the total volume of supernatant in the SC and MC infections. The standard deviations of the $\log_{10}$ total viral titre measurements, $\sigma_\text{RNA,SC}=0.237$ and $\sigma_\text{RNA,MC}=0.258$, were estimated individually for the SC and MC infections, and fixed to the standard deviation of the residuals between the three measurements of $\log_{10}(\Cdata\rna)$ at each sampling time and their mean, pooled over all time points.

\begin{figure}
\begin{center}
\includegraphics[width=\linewidth]{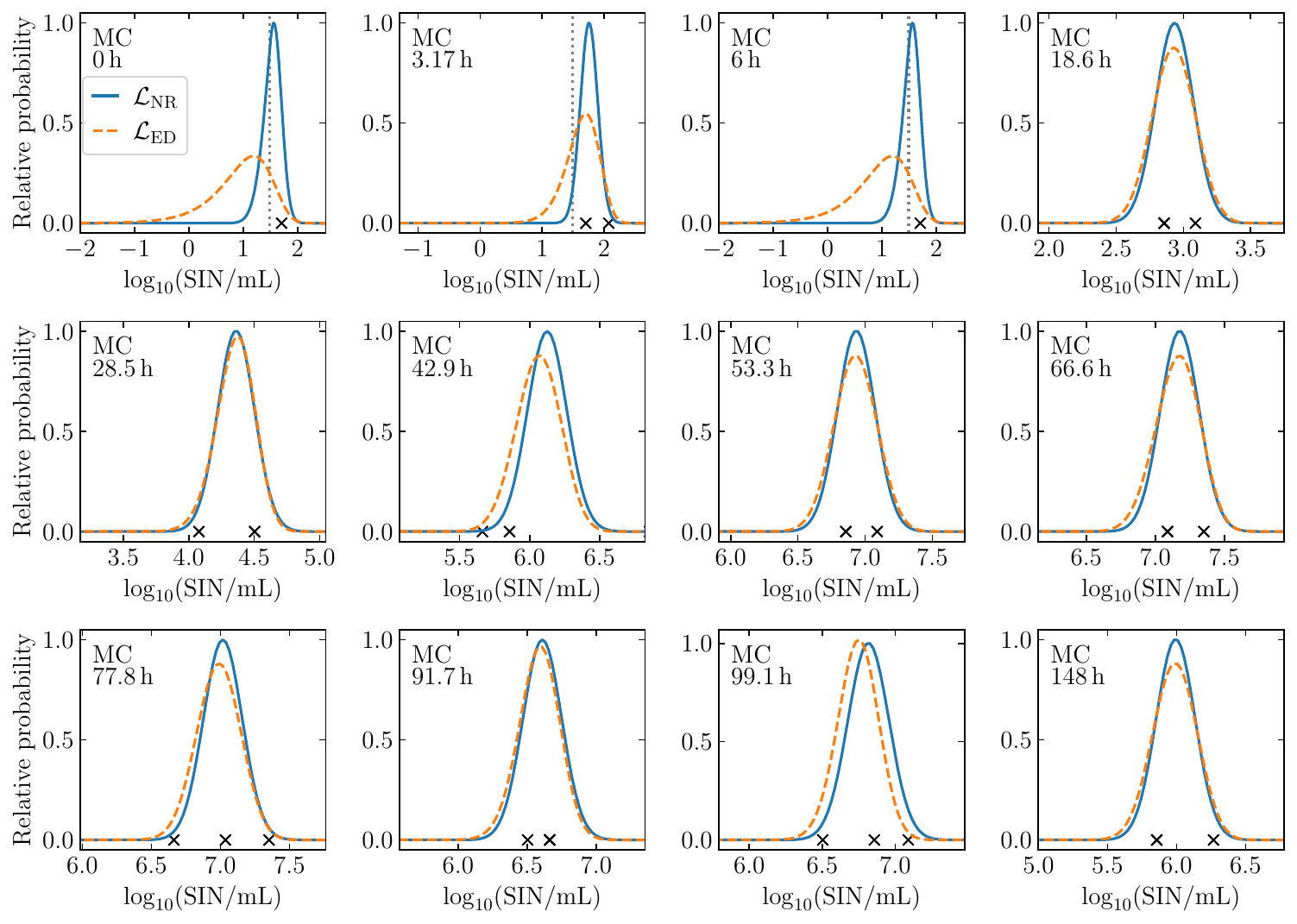}
\caption{%
\textbf{Standard ($\LikeNR$) and alternative ($\LikeED$) likelihood functions over all MC infection samples.}
Comparison of the relative probability ($y$ axis) of the three infectious titres measurements, given the $\log_{10}(\SIN/\milli\litre)$ predicted by the MM ($x$ axis), as computed according to either $\LikeNR$ (blue, solid) or $\LikeED$ (orange, dashed). The relative probability shown is the combined probability for all 3 experimental measurements taken over all sampling times over the course of the MC infections, indicated as black $\times$ along the bottom of each graph. The vertical dashed line for measurements at $\le\unit{6}{\hour}$  corresponds to the lower limit of detection of the ED assay ($C_\text{LLoD}$, Table \ref{tab:fixdpars}).
}
\label{likeliMC}
\end{center}
\end{figure}

\begin{figure}
\begin{center}
\includegraphics[width=\linewidth]{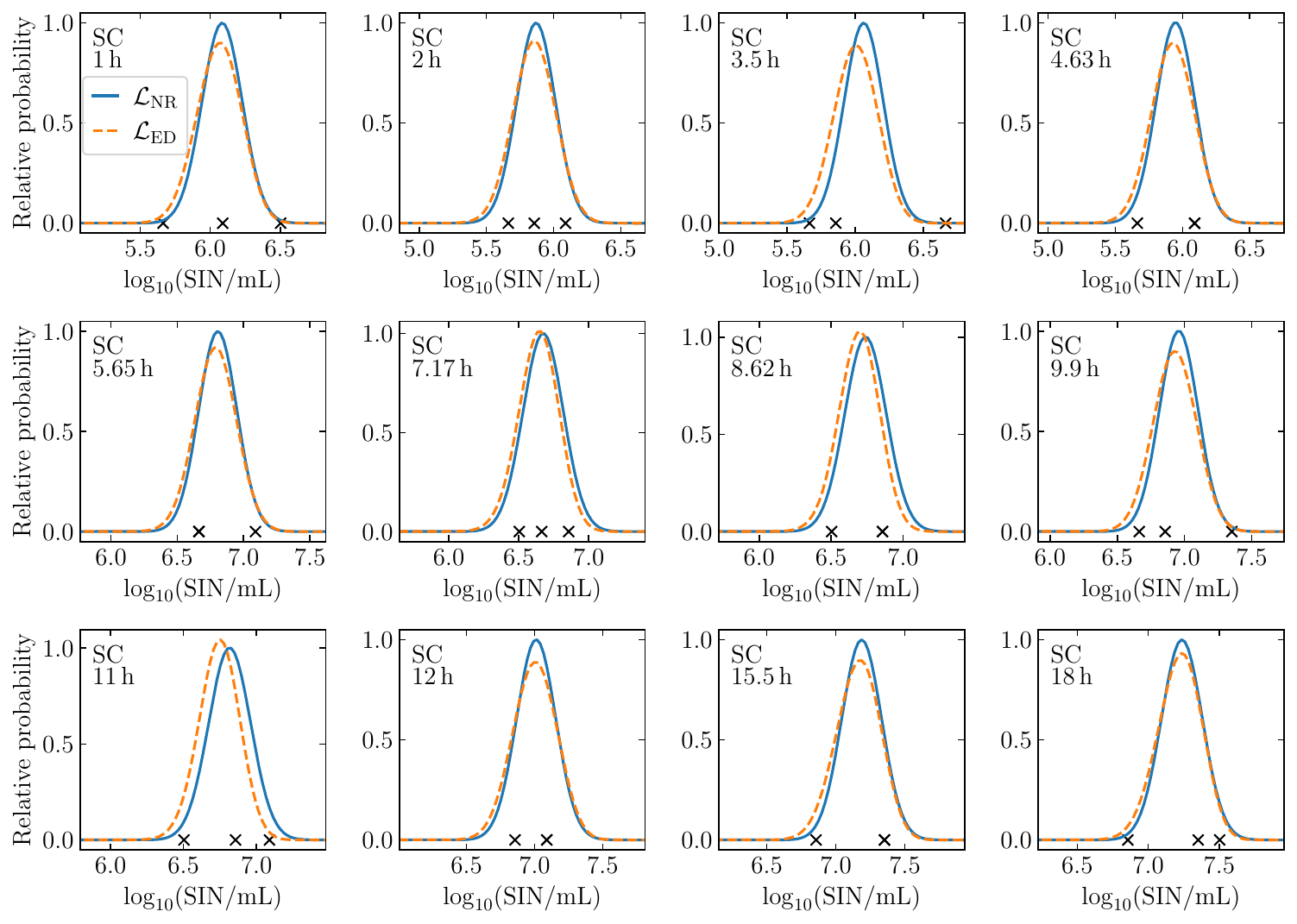}
\caption{%
\textbf{Standard ($\LikeNR$) and alternative ($\LikeED$) likelihood functions over all SC infection samples.}
Comparison of the relative probability ($y$ axis) of the three infectious titres measurements, given the $\log_{10}(\SIN/\milli\litre)$ predicted by the MM ($x$ axis), as computed according to either $\LikeNR$ (blue, solid) or $\LikeED$ (orange, dashed). The relative probability shown is the combined probability for all 3 experimental measurements taken over all sampling times over the course of the SC infections, indicated as black $\times$ along the bottom of each graph.
}
\label{likeliSC}
\end{center}
\end{figure}

For the likelihood of the infectious viral titre measured via an ED assay over the course of the SC and MC infections, $\Like_V(...|...,\pset,\MM)$ in Eqn.\ \eqref{eqn:likepost}, two different distributions were explored. Figures \ref{likeliSC} and \ref{likeliMC} shows these two likelihood functions at each sampling time over the course of the SC and MC infections, computed at each time from the three measurements sampled from replicate infections.

The standard likelihood function, denoted $\LikeNR$, is based on the assumption of normally distributed $\log_{10}(V)$ residuals, as was used for the vRNA measurements, namely
\begin{align}
\Like_V(\Cdata|\vec{\sigma},\pset,\MM) =
\prod_{r=1}^3 \prod_{t=t_\text{SC}}
\LikeNR(\Cdata_{\text{SC},r}(t)|\sigma_\text{SC},\pset,\MM) \cdot 
\prod_{r=1}^3 \prod_{t=t_\text{MC}}
\LikeNR(\Cdata_{\text{MC},r}(t)|\sigma_\text{MC},\pset,\MM)
\end{align}
where $\Cdata$ are the midSIN-estimated $\SIN/\milli\litre$ concentrations of each measured sample, $\LikeNR$ is defined in Eqn.\ \eqref{eqn:likenr}, and the standard deviations of the $\log_{10}$ infectious titre measurements, $\sigma_\text{SC} = 0.247$ and $\sigma_\text{MC}=0.240$, for the SC and MC infections, respectively, were estimated based on $\log_{10}(\Cdata)$ in the same manner as those for the vRNA measurements.

The alternative likelihood for the infectious viral titres, denoted $\LikeED$, is the likelihood of the observed ED outcomes (number of positive wells in each dilution column of the assay) for each virus titre sample, namely
\begin{multline}
\Like_V(\vec{k}^\text{data}|\vec{n}^\text{data},\vec{\Dil}^\text{data},\Vinoc,\pset,\text{MM}) =
\prod_{r=1}^3 \prod_{t=t_\text{SC}}
\LikeED(\vec{k}^\text{data}_{\text{SC},r}(t)|\vec{n}^\text{data}_{\text{SC},r}(t),\vec{\Dil}^\text{data}_{\text{SC},r}(t),\Vinoc,\pset,\MM)\ \times \\
\prod_{r=1}^3 \prod_{t=t_\text{MC}}
\LikeED(\vec{k}^\text{data}_{\text{MC},r}(t)|\vec{n}^\text{data}_{\text{MC},r}(t),\vec{\Dil}^\text{data}_{\text{MC},r}(t),\Vinoc,\pset,\MM)
\end{multline}
where $\{\vec{k}^\text{data},\vec{n}^\text{data},\vec{\Dil}^\text{data}\}$ is the set of all observed ED outcomes, when an outcome is something like $\vec{k}^\text{data}_{\text{SC},r}(t) = (8, 8, 8, 5, 5, 0, ...)$, for each measured sample, and $\LikeED$ is defined in Eqn.\ \eqref{eqn:likeed}.

The ODE model has 6 MM parameters, $\pset_\text{ODE}=\,(\beta,\gamma,\rho,\rho\rna,\tau_E,\tau_I)$, to be estimated from the total and infectious viral titre measurements taken over the SC and MC infections. The SM has one additional parameter: the random number seed (\RNS). Linear uniform priors are assumed for $\tau_E\in[0,\infty)$, $\tau_I\in[0,\infty)$, and $\RNS$. Based on \texttt{numpy}'s \texttt{random} module, RNS must be an integer $\in[0,2^{32})$. A $\log_{10}$ uniform prior was assumed for all other estimated parameters such that
\begin{align}
\prior(\pset) \propto \frac{1}{\gamma \cdot \beta \cdot \rho \cdot \rho\rna}
\label{prior}
\end{align}
where $\rho\rna\in[0,\infty)$, and bounds on the remaining parameters depends on the units of $[V]$. When $[V]=\SIN$, both $\rho$ and $\gamma \in[0,\infty)$, but $\beta\in[\unit{10^{-10}}{\milli\litre/(cell\cdot\hour)},\infty)$ because the data does not inform $\beta$'s lower bound, as explained in the text. When $[V]=\IV$, $\beta\in[0,\infty)$, $\gamma\in\unit{[0,1]}{\IV/cell}$, and $\rho\in[0,\rho\rna\times(\unit{1}{\IV/vRNA})]$ since we cannot produce more $\IV$ than we produce virions. In addition to these bounds on individual parameters, we further constrain $\pset$ by rejecting unphysical outcomes when $[V]=\IV$. Specifically, we require that all initial infectious titers, which are expressed in $\SIN$ in Table \ref{tab:fixdpars} ($V(-\unit{1}{\hour})_\text{SC}$, $V(0)_\text{SC}$, $V(0)_\text{MC}$), once converted to $\IV$ (wherein $V_{\IV}=V_{\SIN}/\PVest$), result in no more $\IV$ than there are virions (vRNA) measured at that time ($V\rna(-\unit{1}{\hour})_\text{SC}$, $V\rna(0)_\text{SC}$, $V\rna(0)_\text{MC}$), i.e.\ $V_{\IV} = V_{\SIN}/\PVest \le V\rna$ or $\PVest \ge V_{\SIN}/V\rna$. Of all infectious and total viral titers pairs, those measured post-rinse in the SC infection impose the highest (and thus the only relevant) lower bound, namely $\PVest\ge(10^{7.01}/10^{7.77}) = 0.174$.

\begin{figure}
\begin{center}
\includegraphics[width=\linewidth]{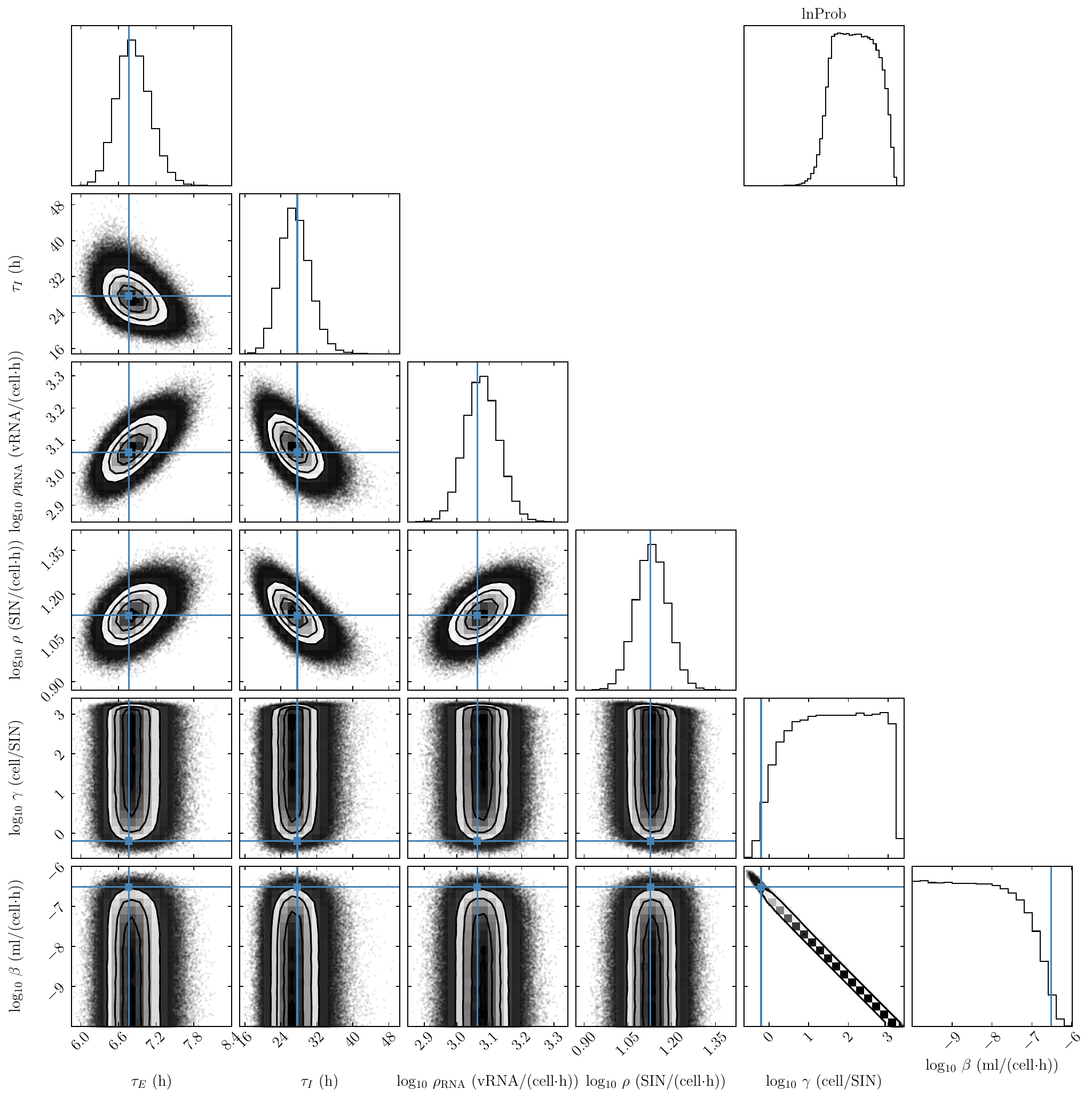}
\caption{\textbf{Pairwise MPDs for the ODE model ($[V]=\SIN$) obtained using the $\LikeNR$ likelihood}.}
\label{tri:mfm_ssr_sin}
\end{center}
\end{figure}

\begin{figure}
\begin{center}
\includegraphics[width=\linewidth]{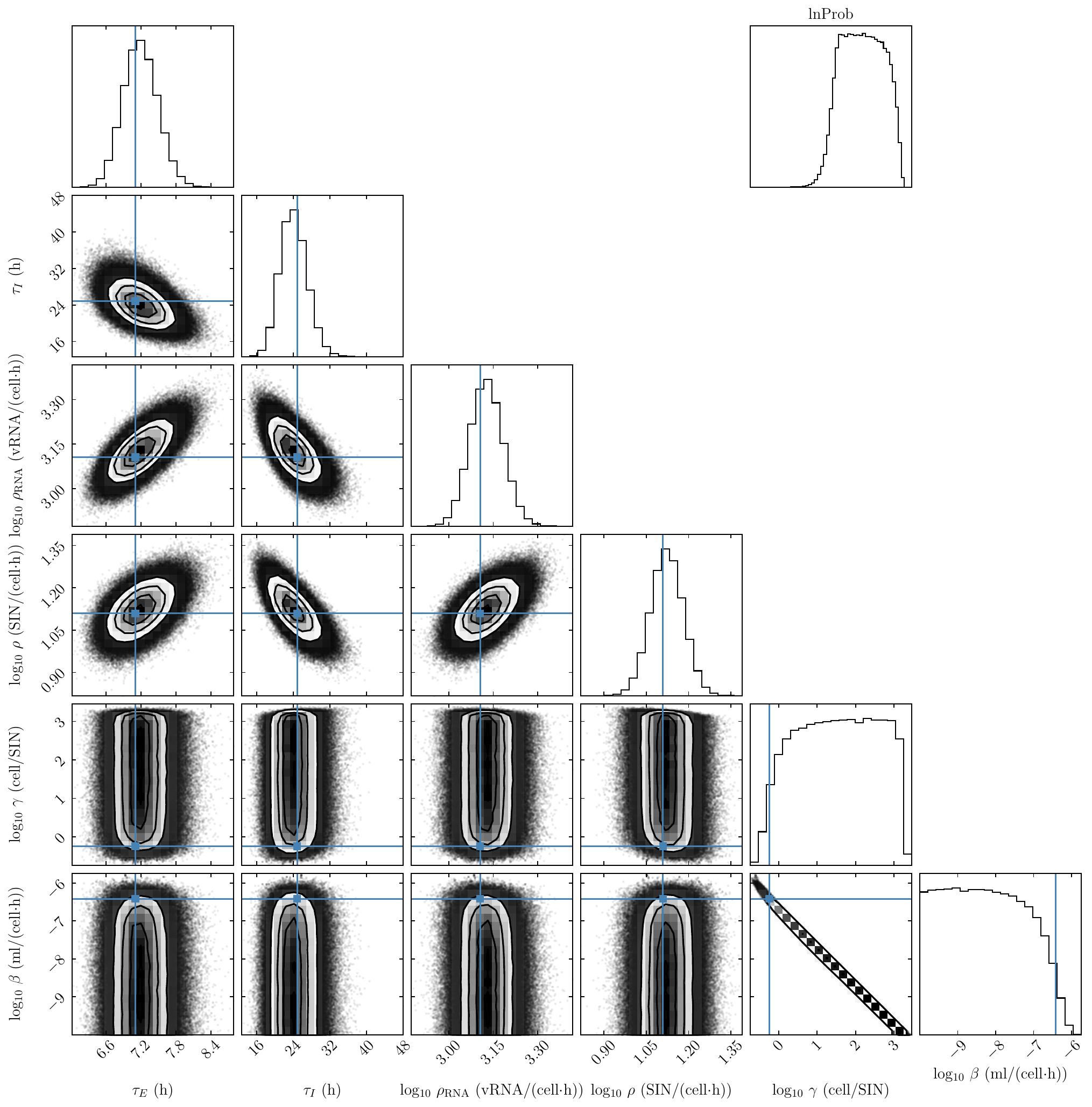}
\caption{\textbf{Pairwise MPDs for the ODE model ($[V]=\SIN$) obtained using the $\LikeED$ likelihood}.}
\label{tri:mfm_ed_sin}
\end{center}
\end{figure}

\begin{figure}
\begin{center}
\includegraphics[width=\linewidth]{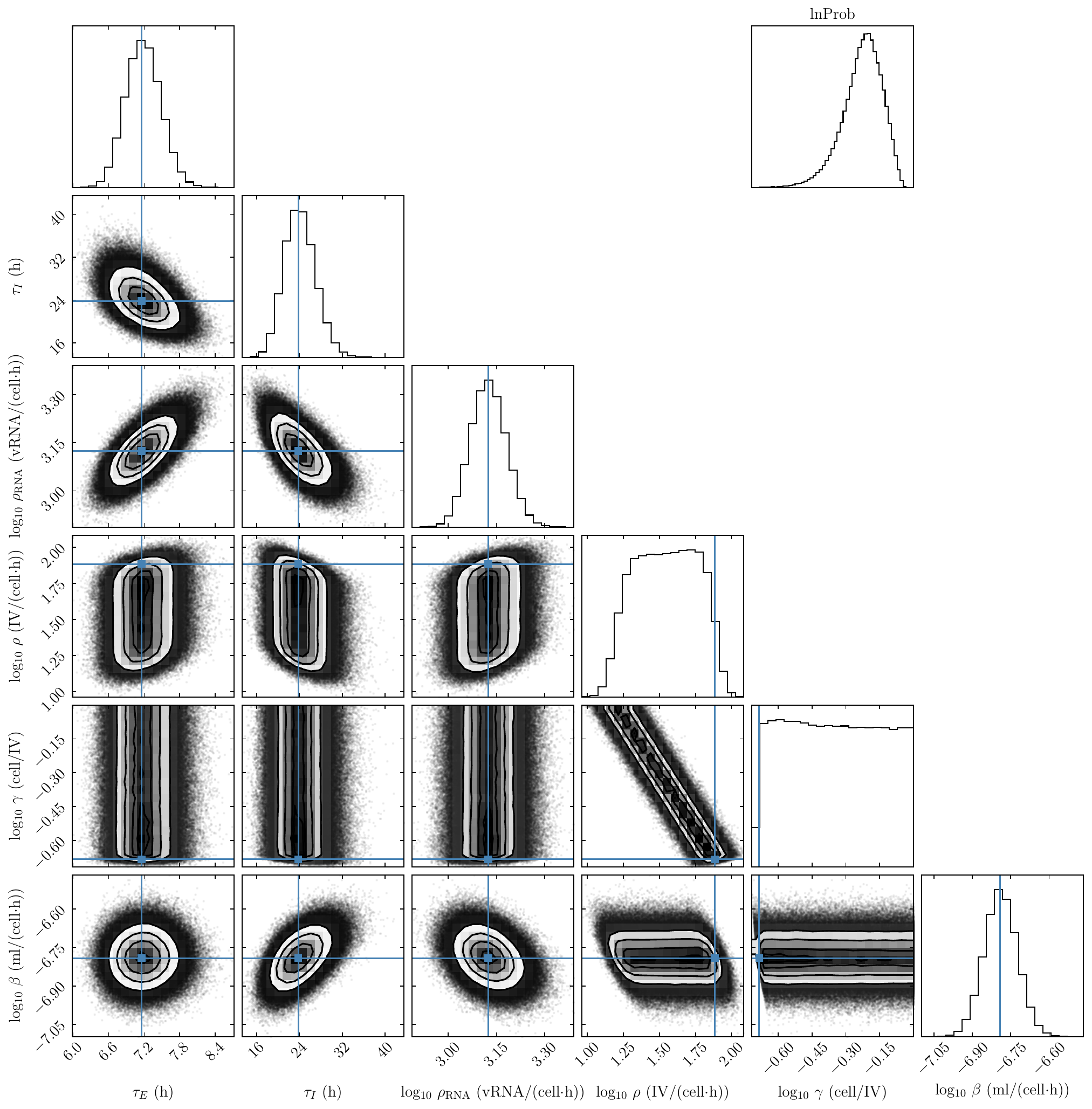}
\caption{\textbf{Pairwise MPDs for the ODE model ($[V]=\IV$) obtained using the $\LikeED$ likelihood}.}
\label{tri:mfm_ed_iv}
\end{center}
\end{figure}

\begin{figure}
\begin{center}
\includegraphics[width=\linewidth]{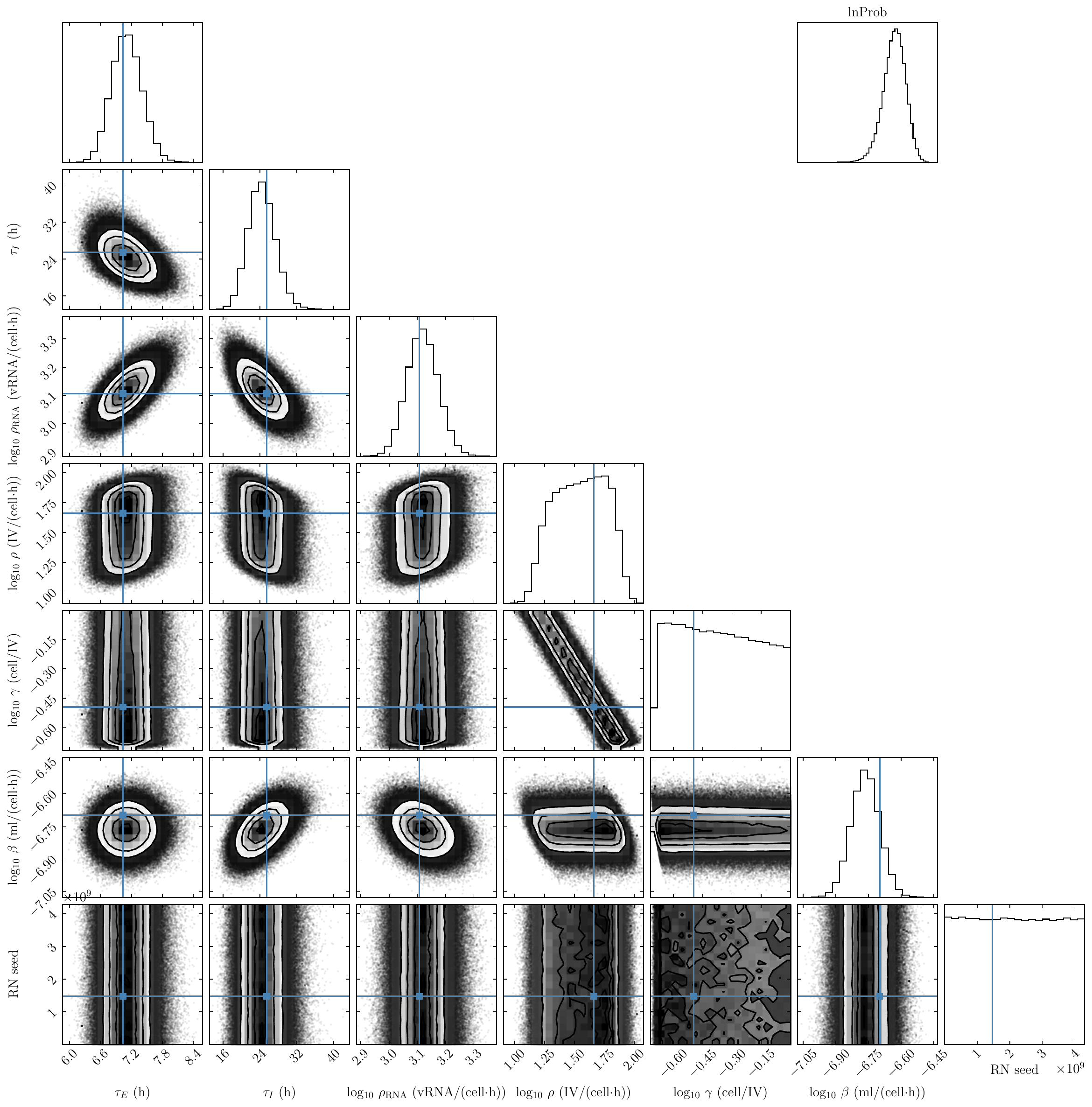}
\caption{\textbf{Pairwise MPDs for the stochastic model ($[V]=\IV$) obtained using the $\LikeED$ likelihood}.}
\label{tri:sm_ed_iv}
\end{center}
\end{figure}

A Markov chain Monte Carlo (MCMC) method, as implemented by \texttt{phymcmc} \cite{phymcmc} which itself relies on the \texttt{emcee} \texttt{Python} module \cite{foreman13}, was used to estimate the parameter posterior distributions (PDs). Given the high degree of correlation between $\gamma$ and $\beta$ when $[V]=\SIN$, parameters ($\gamma$,$\beta$) are replaced with $(a_\SIN=\gamma\cdot\beta,b_\SIN=\gamma/\beta$) in $\pset$ as the parameters to be estimated, such that $\gamma=\sqrt{a_\SIN\cdot b_\SIN}$, $\beta=\sqrt{a_\SIN/b_\SIN}$, and $\prior(\pset) = (a_\text{\SIN}\cdot b_\text{\SIN}\cdot\rho\cdot\rho\rna)^{-1}$. Replacing the two highly correlated parameters ($\gamma$,$\beta$), with two orthogonal quantities ($a_\SIN$,$b_\SIN$) allows for faster convergence because the product $\gamma\cdot\beta$ is well-constrained by the data. Similarly, when $[V]=\IV$, $\gamma$ is highly correlated with $\rho$, and ($\gamma$,$\rho$) are replaced with ($a_\IV=\gamma\cdot\rho,b_\IV=\gamma/\rho$) in $\pset$, such that $\gamma=\sqrt{a_\IV\cdot b_\IV}$, $\rho=\sqrt{a_\IV/b_\IV}$, and $\prior(\pset) = (a_\text{\IV}\cdot b_\text{\IV}\cdot\beta\cdot\rho\rna)^{-1}$. Two-dimensional pairwise marginalized posterior distributions (MPDs) for all MM variants are provided in Figures \ref{tri:mfm_ssr_sin}--\ref{tri:sm_ed_iv}.

The MCMC runs for each variation explored herein were composed of multiple sequential rounds of 300 walkers performing 10\,000 MCMC steps, each started from the end of the previous one, except for the first round which was normally distributed around a reasonable parameter set. The \texttt{emcee}-provided function \texttt{autocorr.integrated\_time} was used to estimate the amount of autocorrelation between successive MCMC steps, which in turn was used to choose the number of steps to discard at the start of the first chain ($\sim$2.5$\times$ the autocorrelation time) to eliminate any residual effect from the chosen initial starting positions, and the thinning factor to apply to each chain (at least half of the autocorrelation time). The final posterior distribution for each variation, after the 2\,000 step burn-in and the thinning, keeping one step out of every $[50,180]$ steps, were applied, comprised 500\,000 steps.

Computationally, each of the sequential MCMC round of 300 chains of 10\,000 steps took $\sim\unit{4.3}{\hour}$ with the ODE model compared to $\sim\unit{22}{\hour}$ with the SM, using all threads of the 36 cores, dual-threaded, \unit{3.5}{\giga\hertz} Intel Xeon 8360Y processors. The autocorrelation time for the ODE with $[V]=\IV$ was the shortest ($\sim$100 steps), followed by the ODE with $[V]=\SIN$ ($\sim$200 steps), with the longest being that for the SM with $[V]=\IV$ ($\sim$300 steps). This means that, in addition to taking $5\times$ longer to accept $10\,000$ steps, $3\times$ more MCMC rounds had to be performed to account for the $3\times$ higher thinning factor applied to chains for the SM. In total, obtaining the final 500\,000 independent parameter sets took $\sim15\times$ longer using the SM (27.5 days) compared to the ODE model (1.79 days).

The 95\% CI for the MM-predicted viral time courses (e.g., panels (A--D) in Figures \ref{fig:varylike}, \ref{fig:varyunits}, \ref{fig:varySM}), were obtained by drawing 10,000 parameter sets at random with replacement from the 500\,000 parameter sets that make up the MCMC-estimated posteriors, and using them to generate 10\,000 time courses. The upper and lower 95\% CI curves shown correspond to the largest and smallest value of $y$ (viral titer) after rejecting the upper and lower 2.5\% of the trajectories at each time ($x$ value).

\subsection{Derivation of the ED assay-based likelihood, $\LikeED$}

The expression for $\LikeED$ in Eqn.\ \eqref{eqn:likeed} herein is derived from Eqn.\ (5) in \cresta which, following the notation therein, gives the probability of observing a number of infected wells in each column of the assay ($\vec{k}_J$) given $q_\text{noinf}(\pset)$, the probability that the well will not become infected given parameter set $\pset$, namely,
\begin{align}
\LikeED(\vec{k}_J|\vec{n}_J,\vec{\Dil}_J,\Vinoc,\pset) = \mathcal{Q}(\vec{k}_J|q_\text{noinf}(\pset)) = \left[ \prod_{j=1}^J\left(1-\left[q_\text{noinf}^i(\pset)\right]^{\mathcal{D}_j}\right)^{k_j^i} \right] \left[q_\text{noinf}^i(\pset)\right]^{\sum_{j=1}^J\mathcal{D}_j\left(n_j^i-k_j^i\right)} \label{cresta5}
\end{align}
where $J$ is the total number of columns, $\mathcal{D}_j$ is the dilution factor in column $j$, $k_j$ is the number of infected wells in column $j$, and $n_j$ is the total number of wells in column $j$. In \cresta, $q_\text{noinf}$ is the probability of observing no infection in an ED well, having drawn a volume $\Vinoc$ of the sample containing $C_\text{measured}$ infection-causing units (SIN) per millilitre into that well. This probability is given by the binomial probability mass function, $\text{Binomial}(k=0|n=\Vinoc/V_\text{vir},p=C_\text{measured}\cdot V_\text{vir})$, evaluated for no success, i.e.\ no infection ($k=0$), where $V_\text{vir}$ is the volume of one virion, $p=C_\text{measured}\cdot V_\text{vir}$, is the probability of success for each draw, and $n=\Vinoc/V_\text{vir}$ is the number of draws.

Here we define $\widetilde{q}_\text{noinf}$ the probability that the well will not become infected given a sample's actual infectious virion concentration $C_\text{actual}$ in $\IV/\milli\litre$. Since it is possible to draw an infectious virion and yet fail to infect the well, we now need to account for additional outcomes: the probability that $k>0$ virions are drawn but all $k$ infectious virions fail to cause an infection, $(\PVext)^k$. Therefore,
\begin{align}
\widetilde{q}_\text{noinf} &= \sum_{k\,=\,0}^n {(\PVext)}^k \cdot \text{Binomial}(k|n\,=\,\Vinoc/V_\text{vir},p\,=\,C_\text{actual}\cdot V_\text{vir}) \nonumber \\
&= \sum_{k\,=\,0}^n {(\PVext)}^k {n \choose k}p^k (1-p)^{n-k} \nonumber \\
&= \sum_{k\,=\,0}^n {n \choose k}(p\cdot \PVext)^k (1-p)^{n-k} \ .
\end{align}
Using the Binomial theorem, the above equation can be written as
\begin{align}
\widetilde{q}_\text{noinf} = [(1-p)+ p \cdot \PVext]^n = [1-p(1-\PVext)]^n = [1-p \cdot \PVest]^{\Vinoc/V_\text{vir}} \ ,
\end{align}
and using the same approximation used in \cresta,
\begin{align}
\text{ln}(1-x)\,&\overset{|x|\ll 1}{\approx}\,-x \nonumber \\
\text{ln}(\widetilde{q}_\text{noinf}) &= \frac{\Vinoc}{V_\text{vir}} \text{ln}(1-p\cdot \PVest) = \frac{\Vinoc}{V_\text{vir}} \text{ln}(1-C_\text{actual}\cdot V_\text{vir}\cdot \PVest) \nonumber \\
\text{ln}(\widetilde{q}_\text{noinf}) &\approx \frac{\Vinoc}{V_\text{vir}} (-C_\text{actual}\cdot V_\text{vir}\cdot \PVest)\,=\,-C_\text{actual}\cdot\Vinoc \cdot \PVest \nonumber \\
\widetilde{q}_\text{noinf} &\approx \exp [ -C_\text{actual}\cdot \PVest \cdot \Vinoc] \label{newqnoinf} \ .
\end{align}
Eqn.\ \eqref{eqn:likeed} replaces ${q}_\text{noinf}(\pset)$ with $\widetilde{q}_\text{noinf}(\pset)$ in Eqn.\ \eqref{cresta5}, and expresses $\widetilde{q}_\text{noinf}(\pset)$ as $\exp[-C^\text{model}(t|\pset) \cdot \Vinoc]$ when the MM expresses $V$ in $\SIN$ or as $\exp[-C^\text{model}(t|\pset) \cdot \PVest \cdot \Vinoc]$ when $[V]=\IV/\milli\litre$, where $\PVest(\pset)$ is the establishment probability of an infection initiated with one infectious virion, given parameter set $\pset$.

\subsection{Lower limit of detection for an ED assay}
\label{LLoD}

Let us consider $\mathcal{L}(\vec{k}_J|q_\text{noinf})$, the probability of observing a number of infected wells per column $\vec{k}_J$ given $q_\text{noinf}$ the probability that the well will not become infected given a sample's measured infection concentration $C_\text{measured}$. This expression is given in \cresta (Eqn.\ (5), $\mathcal{Q}(\vec{k}_J|q_\text{noinf})$, therein) as follows,
\begin{align}
\mathcal{L}(\vec{k}_J|q_\text{noinf}) = \left[ \prod_{j=1}^J (1-q_\text{noinf}^{\mathcal{D}_j})^{k_j} \right] q_\text{noinf}^{\sum_{j=1}^J\mathcal{D}_j(n_j-k_j)}
\end{align}
where $J$ is the total number of columns, $\mathcal{D}_j$ is the dilution factor in column $j$, $k_j$ is the number of infected wells in column $j$, and $n_j$ is the total number of wells in column $j$. The probability of observing no infected wells ($\vec{k}_J = \mathbf{0}$) given $q_\text{noinf}$ is then given by the following expression, namely,
 \begin{align}
\mathcal{L}(\mathbf{0}|q_\text{noinf}) = \left[ \prod_{j=1}^J (1-q_\text{noinf}^{\mathcal{D}_j})^{0} \right] q_\text{noinf}^{\sum_{j=1}^J\mathcal{D}_j(n_j-0)} = q_\text{noinf}^{\sum_{j=1}^J\mathcal{D}_jn_j}
\end{align}

The lower limit of detection could then be defined as the concentration in $\SIN/\milli\litre$ ($C_\text{LLoD}$) at which the probability of observing no infected wells given $q_\text{noinf}$, $\mathcal{L}(\mathbf{0}|q_\text{noinf})$, is 0.5 or 50\%. This can be determined by solving the following equation, and using $q_\text{noinf} = \exp[-C_\text{inf} \cdot \Vinoc]$ (Eqn.\ (2) in \cresta) where $C_\text{inf}$ corresponds to $C_\text{LLoD}$ which we seek to determine, namely
\begin{align}
0.5 &= q_\text{noinf}^{\sum_{j=1}^J \mathcal{D}_jn_j} \nonumber \\
0.5 &= \exp\left[-C_\text{LLoD} \cdot \Vinoc\cdot \sum_{j=1}^J\mathcal{D}_jn_j\right] \nonumber \\
-\ln(0.5) &= C_\text{LLoD} \cdot \Vinoc \cdot \sum_{j=1}^J\mathcal{D}_jn_j \nonumber \\
C_\text{LLoD} &= \frac{-\ln(0.5)}{\Vinoc \cdot \sum_{j=1}^J\mathcal{D}_jn_j}
\end{align}
where $C_\text{LLoD}$ is the ED assay's lower limit of detection in $\SIN/\milli\litre$. Herein, $\Vinoc=\unit{0.05}{\milli\litre}$, $n_j=4\,\forall j$, and $D_j=\left\{10^{-1},10^{-2},...,10^{-8}\right\}$ such that $C_\text{LLoD} = \unit{10^{1.494}}{\SIN/\milli\litre}$.

\subsection{Infecting time in the stochastic model}
\label{tinf-section}

Let us derive the probability that the infecting time ($\tinf$) is of time $t$, i.e.\ the probability that it takes time $t$ for an infectious cell to cause at least one cell to be newly infected, given an uninfected population of cells and neglecting virus loss through loss of infectivity or entry into cells. 

In this case, the relevant SM equations, the equations representing $E_1^{n+1}$ the number of cells in the 1$^\mathrm{st}$ eclipse compartment (i.e.\ newly infected) and $V^{n+1}$ the number of infectious virions, both at time $t = (n+1)\Delta t$, can be simplified to
\begin{align*}
E_1^{n+1} &= E_1^n + N^{\text{inf}} \\
V^{n+1} &= V^n + M^n
\end{align*}
where $N^\text{inf}$ is drawn from $\text{Binomial}(V^n,\, \pinf= \Delta t\cdot \gamma \beta N_\text{cells}/\Vol)$ and $M^n$ is drawn from $\text{Poisson}(\lambda=\Delta t\cdot \rho)$ ($\equiv Q(M^n)$ for convenient notation). The probability that there is no newly infected cell over the next time interval $\Delta t$ is given by $\text{Binomial}(0|V^n,\, \pinf) = (1-\pinf)^{V^n}$. $V^n$ can also be expressed as $\sum_{m=0}^{n-1}M^n$. Therefore, the probability that the infecting time $\tinf = (n+1)\Delta t$ given the set of values $\{M^n\}$, can be expressed as
\begin{align}
\mathcal{P}(t_\text{inf} = (n+1)\Delta t|\{M^n\}) &= \left[ \prod_{m=1}^{n-1}(1-\pinf)^{\sum_{l=0}^{m-1}M^l}\right]\left[1-(1-\pinf)^{\sum_{m=0}^{n-1}M^m}\right] \nonumber \\
&= (1-\pinf)^{\sum_{m=0}^{n-2}(n-1-m)M^m}-(1-\pinf)^{M^{n-1}+\sum_{m=0}^{n-2}(n-m)M^m}
\end{align}

The probability that the infecting time $\tinf = (n+1)\Delta t$ is given by averaging the conditional probability $\mathcal{P}(t_\text{inf} = (n+1)\Delta t|\{M^n\})$ over the distribution $Q(M^n)$ for all values of $\{M^n\}$, namely,
\begin{align}
\mathcal{P}(t_\text{inf} = (n+1)\Delta t) &= \left\langle(1-\pinf)^{\sum_{m=0}^{n-2}(n-1-m)M^m}-(1-\pinf)^{M^{n-1}+\sum_{m=0}^{n-2}(n-m)M^m}\right\rangle_Q \nonumber \\
&= \left\langle(1-\pinf)^{\sum_{m=0}^{n-2}(n-1-m)M^m}\right\rangle_Q-\left\langle(1-\pinf)^{M^{n-1}+\sum_{m=0}^{n-2}(n-m)M^m}\right\rangle_Q \nonumber \\
\end{align}

The random variables $\{M^n\}$ are independent, hence,
\begin{align}
&\mathcal{P}(t_\text{inf} = (n+1)\Delta t) = \prod_{m=0}^{n-2}\left\langle(1-\pinf)^{(n-1-m)M^m}\right\rangle_Q-\left\langle(1-\pinf)^{M^{n-1}}\right\rangle_Q\prod_{m=0}^{n-2}\left\langle(1-\pinf)^{(n-m)M^m}\right\rangle_Q \nonumber \\
&= \prod_{m=0}^{n-2}\left[\me^{-\lambda}\sum_{M^m = 0}^\infty\frac{[(1-\pinf)^{(n-1-m)}\lambda]^{M^m}}{M^m!}\right]-\left[\me^{-\lambda}\sum_{M^{n-1} = 0}^\infty\frac{[(1-\pinf)\lambda]^{M^{n-1}}}{M^{n-1}!}\right]\prod_{m=0}^{n-2}\left[\me^{-\lambda}\sum_{M^m = 0}^\infty\frac{[(1-\pinf)^{(n-m)}\lambda]^{M^m}}{M^m!}\right] \label{tinfexp}
\end{align}

Using the power series representation of the exponential function, i.e.\ $\exp(x) = \sum_{k=0}^\infty x^k/k!$, Eqn.\ \eqref{tinfexp} can be simplified to
\begin{align}
&\mathcal{P}(t_\text{inf} = (n+1)\Delta t) = \prod_{m=0}^{n-2}\exp\left([(1-\pinf)^{(n-1-m)}-1]\lambda\right)-\exp\left([(1-\pinf)-1]\lambda\right)\prod_{m=0}^{n-2}\exp\left([(1-\pinf)^{(n-m)}-1]\lambda\right) \nonumber \\
&= \exp\left(-(n-1)\lambda+(1-\pinf)^{n-1}\lambda\sum_{m=0}^{n-2}(1-\pinf)^{-m}\right)-\exp\left(-\pinf\lambda-(n-1)\lambda+(1-\pinf)^n\lambda\sum_{m=0}^{n-2}(1-\pinf)^{-m}\right) \label{tinfgeo}
\end{align}

The partial sums are of geometric series, therefore, Eqn.\ \eqref{tinfgeo} can be evaluated to
\begin{align}
&\mathcal{P}(t_\text{inf}= (n+1)\Delta t) \nonumber \\
&= \exp\left(\frac{(1-\pinf)-(1-\pinf)^n}{\pinf}\lambda-(n-1)\lambda\right)\left[1-\exp\left(-\pinf\lambda-(1-\pinf)\lambda+(1-\pinf)^n\lambda\right)\right] 
\end{align}

To obtain the probability density function for the infecting time $\tinf$ from the above probability mass function, we need to take the following limit, 
\begin{align}
&\mathcal{P}(t_\text{inf}=t)
= \lim_{\Delta t \to 0}\frac{\mathcal{P}(t_\text{inf}= (n+1)\Delta t)}{\Delta t} \nonumber \\
&= \lim_{n \to \infty} \exp\left(\frac{\rho}{\alpha}\left[1-\frac{\alpha t}{n}-\left(1-\frac{\alpha t}{n}\right)^n\right]-\rho t+\frac{\rho t}{n}\right)\left[1-\exp\left(-\frac{\alpha \rho t^2}{n^2}-\frac{\rho t}{n}\left[1-\frac{\alpha t}{n}-\left(1-\frac{\alpha t}{n}\right)^n\right]\right)\right] \frac{n}{t} \label{limit1}
\end{align}
where $\alpha = \gamma \beta N_\text{cells}/\Vol$. As the exponent of the exponential in the second term of Eqn.\ \eqref{limit1} is small as $n$ approaches $\infty$, we can use the Taylor series approximation $\me^x \approx 1+x$ to approximate this term, namely,
\begin{align}
\mathcal{P}(t_\text{inf}=t)
&= \lim_{n \to \infty} \exp\left(\frac{\rho}{\alpha}\left[1-\frac{\alpha t}{n}-\left(1-\frac{\alpha t}{n}\right)^n\right]-\rho t+\frac{\rho t}{n}\right)\left(\frac{\alpha \rho t^2}{n}+\rho\left[1-\frac{\alpha t}{n}-\left(1-\frac{\alpha t}{n}\right)^n\right]\right) \nonumber \\ 
&= \rho \left[1-\me^{-\alpha t}\right]\exp\left(\frac{\rho }{\alpha}\left[1-\me^{-\alpha t}\right]-\rho t\right) \label{tinfpdfshort}
\end{align}
Substituting $\alpha = \gamma \beta N_\text{cells}/\Vol$ in Eqn.\ \eqref{tinfpdfshort}, we have 
\begin{align}
\mathcal{P}(t_\text{inf}=t) &= \rho\left[1-\me^{-\gamma \beta N_\text{cells}/\Vol \cdot t}\right]\exp\left(\frac{\rho}{\gamma \beta N_\text{cells}/\Vol}\left[1-\me^{-\gamma \beta N_\text{cells}/\Vol \cdot t}\right]-\rho t\right) \label{tinfpdf}
\end{align}

Let us now derive the mean infecting time using Eqn.\ \eqref{tinfpdf},
\begin{align}
\langle \tinf \rangle &= \int_0^\infty t \cdot \mathcal{P}(t_\text{inf}=t)\mathrm{d}t \nonumber \\
&= \int_0^\infty \rho\,t (1-\me^{-\alpha t}) \cdot \me^{\frac{\rho}{\alpha} (1-\me^{-\alpha t})-\rho\,t}\mathrm{d}t \nonumber \\
&= -\frac{\rho}{\alpha^2}\me^{\frac{\rho}{\alpha}}\int_0^1 \ln(y) \cdot (1-y) \cdot \me^{-\frac{\rho}{\alpha} y} \cdot y^{\frac{\rho}{\alpha}-1}\mathrm{d}y \nonumber \\
&= \frac{1}{\alpha}\me^{\frac{\rho}{\alpha}}\int_0^1\me^{-\frac{\rho}{\alpha}\left[y-\left(1-\frac{\alpha}{\rho}\right)\ln(y)\right]}\mathrm{d}y \label{prelaplace}
\end{align}
where $y = \me^{-\alpha t}$. Let us then use Laplace's method to approximate the above integral, assuming $\rho/\alpha$ is a large number. To start, the function in the exponent $f(y) = -(\rho/\alpha)\left[y-\left(1-\alpha/\rho\right)\ln(y)\right]$ can be written as a Taylor series approximation around $y_0 = 1-\alpha/\rho$ where $f'(y_0) = 0$, namely,
\begin{align}
f(y) \approx -\frac{\rho}{\alpha}\left(1-\frac{\alpha}{\rho}-\left(1-\frac{\alpha}{\rho}\right)\ln\left(1-\frac{\alpha}{\rho}\right)+\frac{1}{2}\frac{1}{1-\frac{\alpha}{\rho}}\left[y-\left(1-\frac{\alpha}{\rho}\right)\right]^2\right)
\end{align}
and Eqn.\ \eqref{prelaplace} can be approximated as
\begin{align} 
\langle \tinf \rangle &\approx \frac{1}{\alpha}\me^{\frac{\rho}{\alpha}\left[\frac{\alpha}{\rho}+\left(1-\frac{\alpha}{\rho}\right)\ln\left(1-\frac{\alpha}{\rho}\right)\right]} \int_0^1 \me^{-\frac{1}{2}\frac{\rho}{\alpha}\frac{1}{1-\frac{\alpha}{\rho}}\left[y-\left(1-\frac{\alpha}{\rho}\right)\right]^2}\mathrm{d}y \label{laplace}
\end{align}
As $\rho/\alpha$ approaches $\infty$, Eqn.\ \eqref{laplace} becomes
\begin{align}
\langle \tinf \rangle &\approx \frac{1}{\alpha} \cdot \frac{1}{2}\cdot \sqrt{\frac{2\pi}{\rho/\alpha}} \nonumber \\
&\approx \sqrt{\frac{\pi}{2 \rho \cdot \alpha}} = \sqrt{\frac{\pi}{2 \rho \cdot \gamma \beta N_\text{cells}/\Vol}} \label{avgtinf}
\end{align}
where here we have used the fact that the interval $[0,1]$ covers approximately half of the neighbourhood of $y_0 = 1-\alpha/\rho \approx 1$.

\begin{figure}
\begin{center}
\includegraphics[width=\linewidth]{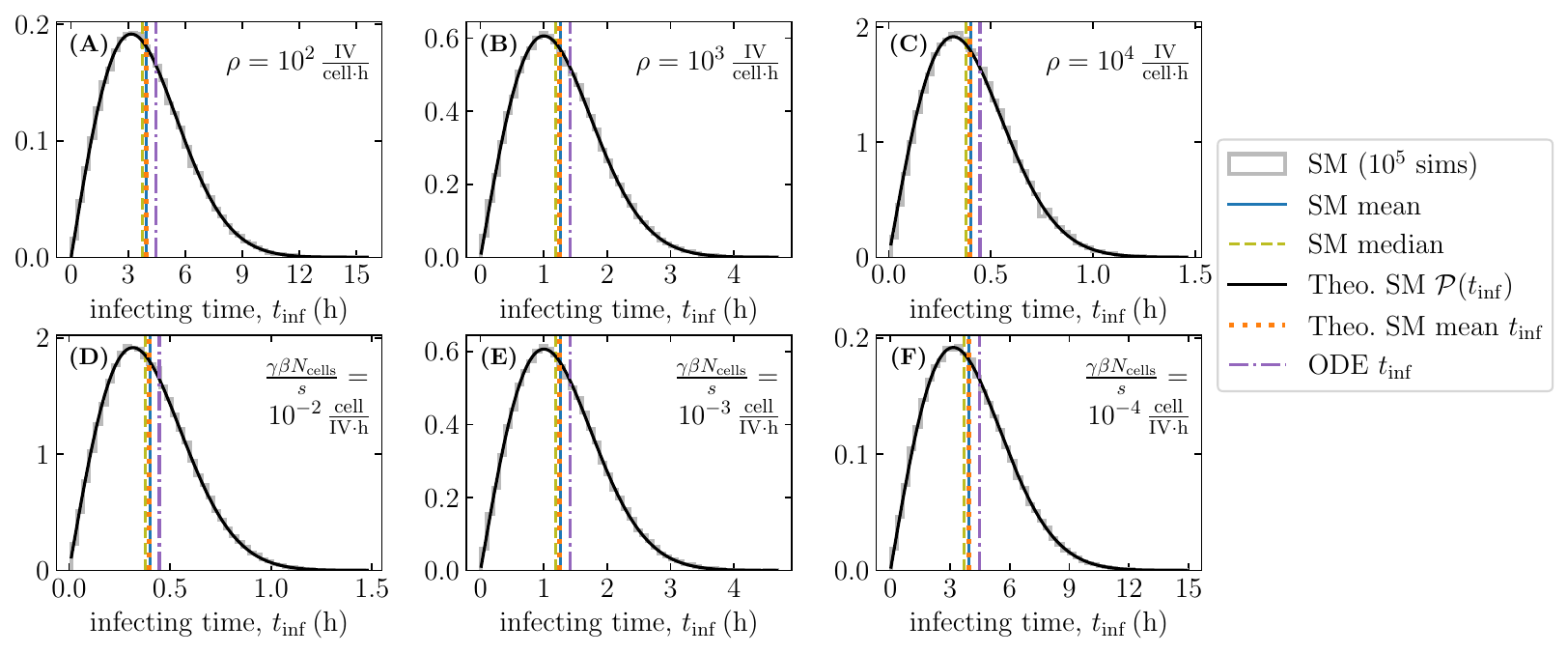}
\caption{%
\textbf{Infecting time distribution in the stochastic model.} %
Density-normalized histograms for the infecting time of $10^5$ SM simulations, compared to the theoretically predicted distribution, $\mathcal{P}(\tinf)$, Eqn.\ \eqref{tinfpdf}, for different values of (A,B,C) $\rho$ or (D,E,F) $\gamma \beta N_\text{cells}/\Vol$. The mean and median of the $10^5$ SM simulations are shown, along with the theoretically predicted SM mean of $\tinf$, Eqn.\ \eqref{avgtinf}. The SM-simulated infecting time corresponds to the duration of SM simulations initiated with a single infectious cell ($I_1(0)=\unit{1}{cell}$) that are terminated when at least one cell has become infected ($E_1(t_\text{inf}) \geq \unit{1}{cell}$), in the absence of any virus loss ($c = \unit{0}{h^{-1}}$, $\gamma = \unit{1}{cell/IV}$). Unless otherwise stated, $\rho = \unit{10^3}{\IV/(cell \cdot\hour)}$ and $\gamma \beta N_\text{cells}/\Vol = \unit{10^{-3}}{cell/(\IV \cdot\hour)}$. Though they do not have an effect on the infecting time, the other infection parameters were $n_E = n_I = 60$, $\tau_E = \unit{7}{\hour}$, $\tau_I = \unit{41}{\hour}$, $N_\text{cells} = \unit{1.9 \times 10^6}{cells}$, $\Vol = \unit{1}{\milli\litre}$, and $\rho_\text{RNA} = c_\text{RNA} = 0$.
}
\label{tinf}
\end{center}
\end{figure}

Figure \ref{tinf} shows that the normalized histogram of the infecting time, generated from $10^5$ SM simulations, is in agreement with Eqn.\ \eqref{tinfpdf}, over a wide range of infection parameters.

%%%%%%%%%%%%%%%%%%%%%%%%%%%%%%%%%%%%%%%%%%%%%%%%%%%%%%%%%%%%%%%%%%%%%%%%%%%%%%%%
%%%%%%%%%%%%%%%%%%%%%%%%%%%%%%%%%%%%%%%%%%%%%%%%%%%%%%%%%%%%%%%%%%%%%%%%%%%%%%%%
%%%%%%%%%%%%%%%%%%%%%%%%%%%%%%%%%%%%%%%%%%%%%%%%%%%%%%%%%%%%%%%%%%%%%%%%%%%%%%%%

\begin{acknowledgements}
This work was supported in part by Discovery Grants RGPIN/3774-2022 (C.A.A.B.) from the Natural Sciences and Engineering Research Council of Canada (NSERC, \url{www.nserc-crsng.gc.ca}), by the Interdisciplinary Theoretical and Mathematical Sciences programme (iTHEMS, \url{ithems.riken.jp}) at RIKEN (C.A.A.B.), and by the Japanese Society for the Promotion of Science (JSPS, \url{www.jsps.go.jp}) KAKENHI Grant Numbers 21H04595 (C.A.A.B.) and 20K14435 (K.A.).
\end{acknowledgements}

\section*{Author Contributions}
C.Q., R.T.\ and C.A.A.B.\ conceptualized the work and wrote the original draft; All authors contributed to parts or all of the investigation, formal analysis, and to the review and editing of the manuscript.

%%%%%%%%%%%%%%%%%%%%%%%%%%%%%%%%%%%%%%%%%%%%%%%%%%%%%%%%%%%%%%%%%%%%%%%%%%%%%%%%
%%%%%%%%%%%%%%%%%%%%%%%%%%%%%%%%%%%%%%%%%%%%%%%%%%%%%%%%%%%%%%%%%%%%%%%%%%%%%%%%
%%%%%%%%%%%%%%%%%%%%%%%%%%%%%%%%%%%%%%%%%%%%%%%%%%%%%%%%%%%%%%%%%%%%%%%%%%%%%%%%

\bibliography{flu-stochastic-2}

%%%%%%%%%%%%%%%%%%%%%%%%%%%%%%%%%%%%%%%%%%%%%%%%%%%%%%%%%%%%%%%%%%%%%%%%%%%%%%%%
%%%%%%%%%%%%%%%%%%%%%%%%%%%%%%%%%%%%%%%%%%%%%%%%%%%%%%%%%%%%%%%%%%%%%%%%%%%%%%%%
%%%%%%%%%%%%%%%%%%%%%%%%%%%%%%%%%%%%%%%%%%%%%%%%%%%%%%%%%%%%%%%%%%%%%%%%%%%%%%%%

\end{document}